\documentclass[aps,prb,twocolumn,10pt,amsmath,amssymb,floatfix,
               nofootinbib,superscriptaddress]{revtex4-2}

\usepackage{graphicx}
\usepackage{bm}
\usepackage{amsthm}
\usepackage{mathtools}
\usepackage{esint}
\usepackage{xcolor}
\usepackage[colorlinks=true,linkcolor=blue!55!black,citecolor=blue!55!black,
            urlcolor=blue!55!black]{hyperref}
\hypersetup{pdftitle={Hidden symmetry at the diabolical points of a biaxial spin},
            pdfauthor={Shadan Ghassemi Tabrizi}}

\theoremstyle{plain}
\newtheorem{result}{Result}

\newcommand{\bh}{\mathbf h}
\newcommand{\bS}{\mathbf S}
\newcommand{\HS}{\mathcal H}
\newcommand{\Ran}{\operatorname{Ran}}
\newcommand{\rank}{\operatorname{rank}}

\newcommand{\sgn}{\operatorname{sgn}}

\newcommand{\ket}[1]{\lvert #1\rangle}
\newcommand{\bra}[1]{\langle #1\rvert}
\newcommand{\ip}[2]{\langle #1 | #2\rangle}
\newcommand{\me}[3]{\langle #1 | #2 | #3\rangle}
\newcommand{\Zp}{Z_{+}}
\newcommand{\Zm}{Z_{-}}
\newcommand{\ii}{\mathrm{i}}
\newcommand{\one}{\mathbf{1}}
\newcommand{\Hpair}{\mathcal{H}_{\mathrm{pair}}}

\begin{document}

\title{Hidden symmetry at the diabolical points of a biaxial spin}

\author{Shadan Ghassemi Tabrizi}
\email{s.ghassemi-tabrizi@hzdr.de}
\affiliation{Computational System Sciences, Technische Universit\"at Dresden,
01187 Dresden, Germany}
\affiliation{Center for Advanced Systems Understanding (CASUS),
Am Untermarkt 20, 02826 G\"orlitz, Germany}

\date{November 11, 2026}

\begin{abstract}
In a rotated frame the biaxial spin Hamiltonian $k_1S_x^2+k_2S_y^2-\bh\cdot\bS$
is a finite tight-binding chain whose hopping amplitudes are tuned by the
applied field. A chain with no vanishing hopping has a nondegenerate spectrum,
so a degeneracy can occur only where the field severs the chain. At every point
of the diabolical-point lattice proven exact by Ke\c{c}ecio\u{g}lu and Garg, the chain is severed in two different rotated frames, at two bonds that are
labeled independently. We write the two severings as two spectral-cut
projectors. These are explicit polynomials in two field-independent spin
components; they commute with the Hamiltonian at that point but not with each
other. This gives a concrete form to the hidden symmetry anticipated by Garg: on the
degenerate subspace the two projectors generate a pseudospin SU(2) that
commutes with the Hamiltonian. The rank of a pairing operator built from them
gives the multiplicity of every lattice point, without the continuity and
topological argument used previously. Because the two partners of a doublet
occupy disjoint stretches of the chain, the splitting determinant factorizes
exactly and has a fixed negative sign. Every degeneracy of the model is therefore
a linear cone, all cones have the same orientation, and the lower level
carries Chern charge $-1$ for the Berry connection $\ii\langle\psi|\nabla\psi\rangle$.
\end{abstract}

\maketitle

% =====================================================================
\section{Introduction}
\label{sec:intro}
% =====================================================================

At low temperature, a molecular magnet such as Fe$_8$ behaves as a single
large spin with strong easy-axis anisotropy and two inequivalent transverse
axes. The tunnel splitting between its two lowest states vanishes at a discrete
set of applied fields. This quenching was predicted by Garg~\cite{Garg1993} as
Berry-phase interference between two tunneling paths, and it was observed in
the oscillations of the tunnel splitting of Fe$_8$ measured by Wernsdorfer and
Sessoli~\cite{WernsdorferSessoli}; quenching of tunnel splittings between
excited levels has more recently been observed in a half-integer-spin
lanthanide magnet~\cite{PaulGd}. The quenching points are diabolical points (DPs), degeneracies of one pair of
levels at isolated points of the three-dimensional space of field
components~\cite{vNW,BerryWilkinson}. Each DP is a Berry-curvature monopole of
integer charge, of unit modulus when the two energy surfaces meet in a
nondegenerate linear cone~\cite{Berry1984,GargAJP}.

For the quadratic biaxial model of a spin $J$,
\begin{equation}
  H(\bh)=k_1S_x^2+k_2S_y^2-\bh\cdot\bS ,
  \qquad k_1>k_2>0 ,
  \label{eq:H}
\end{equation}
the DP structure is known exactly and is remarkably regular. The centered rectangular lattice of DPs in the $h_x$--$h_z$ plane was obtained
semiclassically, with the discrete WKB method, by Garg~\cite{GargI,GargII} and
independently, with an approximate version of that method, by Villain and
Fort~\cite{VillainFort}; both noted that the semiclassical formulas appear to
be exact~\cite{KG}. Ke\c{c}ecio\u{g}lu and Garg~\cite{KG} proved that the
lattice is exact and computed the multiplicity of every lattice point, that is,
the number of level pairs that become degenerate simultaneously at one and the
same field. The quenches that Wernsdorfer and Sessoli also observed at nonzero $h_z$,
between levels 2 and 3 and between levels 3 and 4, are off-axis points, except
for the quench of levels 2 and 3 at $h_x=0$, which lies on the $h_z$ axis; the
oscillation of the splitting with $h_x$ is shifted by half a period for each
excited level of the deeper well~\cite{WernsdorferSessoli}.
Reference~\cite{Garg2000} established off-axis points at finite $J$ by
perturbation theory in $k_2/k_1$, found a lattice point of
multiplicity two, and noted that the off-axis points are ``not associated with
any obvious symmetry of $H$''. Summed over the lattice, the multiplicities give
$\tfrac23J(J+1)(2J+1)$ degenerate level pairs. Bruno subsequently derived
topological sum rules for magnetic DPs and stated that this lattice consists
entirely of ordinary two-level DPs of diabolicity index unity~\cite{Bruno}.

No symmetry of the family $H(\bh)$ had been found that accounts for the
lattice as a whole. On the
$h_x$ and $h_z$ axes a $\pi$ rotation about the field commutes with
\eqref{eq:H} and sorts the states into two symmetry sectors, so that a crossing
between them is not forbidden, but this symmetry does not say at which fields
the crossings occur, and off the axes there is no obvious unitary symmetry that
separates the crossing levels into different sectors. Ke\c{c}ecio\u{g}lu and
Garg obtained the multiplicities by deforming the anisotropy continuously to a
solvable limit and transporting the count topologically, and they wrote that
they ``have not been able to find if the Hamiltonian (1) has a higher symmetry
at the diabolical points''~\cite{KG}. Garg noted that the simultaneity of the
degeneracies and the exactness of the quenching fields point to a higher
dynamical symmetry~\cite{GargI}, and returned to the question later: ``it is
tempting to speculate that the Hamiltonian possesses an additional `hidden'
symmetry, but that is so far unproven''~\cite{GargAJP}.

For one field line the answer is partly known. In an unpublished preprint, Preda
and Barnes~\cite{PredaBarnes} exhibited, for integer spin and the field along
the hard axis, an antilinear operator, time reversal times an odd polynomial in
a complex spin combination, that commutes with $H$ at a periodic sequence of
fields; a $\pi$ rotation about $y$ in place of time reversal gives a linear
version. In the notation of Sec.~\ref{sec:chain} their fields are the points
$h_x=k_1s(2m+1)$ of the lattice studied here on the line $h_z=0$. At the smallest positive one, $h_x=k_1s$, they find $J$ doublets, the multiplicity $f(0,-1)=J$ of
Sec.~\ref{sec:counting}, and their operator, like the pairing operator
\eqref{eq:B}, either annihilates a state or maps it to a degenerate partner.
Their construction is confined to that line and to integer $J$, and it gives
neither the general multiplicity nor the orientation of the cones.

\subsection{The mechanism}

Rotate the spin frame about $y$ so that the quantization axis becomes
\begin{equation}
  Z_\pm=\pm sS_x+cS_z,
  \qquad c=\sqrt{k_2/k_1},\quad s=\sqrt{1-k_2/k_1} .
  \label{eq:Zpm}
\end{equation}
In the eigenbasis of either $Z_+$ or $Z_-$ the full Hamiltonian \eqref{eq:H},
anisotropy and Zeeman term together, is tridiagonal
(Appendix~\ref{app:chain}), that is, a one-dimensional tight-binding chain: the
$N=2J+1$ states $\ket{r}$, $r=-J,\dots,J$, are sites, the diagonal entries are
on-site energies, and the off-diagonal entries are hopping amplitudes, all of
them tuned by the applied field.

A finite chain with every hopping nonzero has a simple spectrum
(Appendix~\ref{app:chain}). A degeneracy therefore requires a hopping to vanish, which severs the chain
into two decoupled pieces whose levels are free to cross. Imposing a vanishing
hopping in the $Z_+$ frame gives one family of straight lines in the field
plane, labeled by an index $m$ running over $-J,\dots,J-1$, and imposing it in
the $Z_-$ frame gives a second family, labeled by $n$ in the same range. The
intersections are the Ke\c{c}ecio\u{g}lu--Garg lattice.

\begin{figure}[tbp]
\centering
\includegraphics[width=\columnwidth]{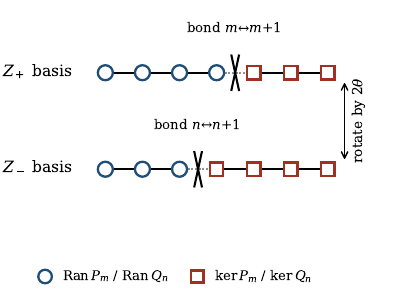}
\caption{The mechanism, for $J=3$ and $(m,n)=(0,-1)$. At a lattice point the
Hamiltonian is a chain of $N=2J+1$ sites severed at the bond
$m\leftrightarrow m+1$ in the eigenbasis of $\Zp$, and at the independently
labeled bond $n\leftrightarrow n+1$ in the eigenbasis of $\Zm$, obtained by
rotating the frame through $2\theta$.}
\label{fig:chains}
\end{figure}

At an intersection the chain is severed in the $Z_+$ frame and in the $Z_-$
frame at once, at bonds that are labeled independently, the two frames being
related by a rotation through $2\theta$ with $c=\cos\theta$
(Fig.~\ref{fig:chains}). Each severing supplies a projector onto the lower part
of its chain,
\begin{equation}
  P_m=\one_{[-J,m]}(\Zp),\qquad Q_n=\one_{[-J,n]}(\Zm) ,
  \label{eq:PQ}
\end{equation}
which commutes with the Hamiltonian. The two projectors do not commute with
each other (Result~\ref{res:transversal} below), and this noncommutation forces
the degeneracies.

\subsection{Relation to previous work}

Table~\ref{tab:novelty} separates what is known about this lattice from what
the present paper adds. Ke\c{c}ecio\u{g}lu and Garg~\cite{KG} proved that the lattice is exact and
computed its multiplicities, locating the lattice with the same two rotated
frames and the same tight-binding reading of the tridiagonal form;
Sec.~\ref{sec:chain} rederives the lattice in the notation used below. New are
(i) the two cut projectors, which commute with $H$ at the lattice point but not
with each other, and their reading as a hidden symmetry with a pseudospin on
the degenerate subspace (Secs.~\ref{sec:cuts} and~\ref{sec:counting}); (ii)
the derivation of the multiplicities from the rank of a pairing operator, at
the physical anisotropy (Result~\ref{res:mult}); and (iii) an exact factorized formula
for the splitting determinant, which proves that every degeneracy is a linear
cone and that all cones have the same orientation (Result~\ref{res:chern}).
With the charge of every DP thus known, Bruno's sum rules~\cite{Bruno},
specialized to this model, follow by counting, which gives an alternative derivation of these known relations.
The operator \eqref{eq:heunform} that $H$ becomes at a lattice point is a known
algebraic Heun operator~\cite{Perline,GVZ,CVZ}.

\begin{table}[tb]
\caption{Relation to previous work. The last column gives the Result or
section of this paper in which the item is treated.}
\label{tab:novelty}
\begin{ruledtabular}
\begin{tabular}{@{}p{0.235\columnwidth}p{0.41\columnwidth}p{0.265\columnwidth}@{}}
\raggedright Item & \raggedright Previous work & \raggedright This work
\tabularnewline
\hline
\raggedright DP lattice, Eq.~\eqref{eq:lattice} &
\raggedright semiclassical~\cite{GargI,GargII,VillainFort}; exact, from two
rotated frames~\cite{KG} &
\raggedright rederived, Sec.~\ref{sec:chain} \tabularnewline
\raggedright Block decomposition at a lattice point &
\raggedright one rotated frame at a time~\cite{KG} &
\raggedright two cut projectors, Result~\ref{res:cuts} \tabularnewline
\raggedright Symmetry at the lattice points &
\raggedright antilinear or linear, on the hard axis and for integer
$J$ only~\cite{PredaBarnes} &
\raggedright noncommuting cuts, pseudo\-spin, Result~\ref{res:transversal},
Eq.~\eqref{eq:factorization} \tabularnewline
\raggedright Multiplicity $f(m,n)$ &
\raggedright deformation to a solvable limit, continuity and
topology~\cite{KG} &
\raggedright rank of $B_{mn}$, Result~\ref{res:mult}; chain count,
App.~\ref{app:levels} \tabularnewline
\raggedright Operator \eqref{eq:heunform} &
\raggedright algebraic Heun operator~\cite{Perline,GVZ,CVZ} &
\raggedright its locking conditions give the lattice, Sec.~\ref{sec:cuts} \tabularnewline
\raggedright Linear cones, unit charge &
\raggedright presupposed~\cite{KG,LiGarg}; stated without
proof~\cite{Bruno,GargAJP} &
\raggedright $\det M\neq0$, Result~\ref{res:chern} \tabularnewline
\raggedright Common orientation &
\raggedright stated, and noted to satisfy the sum rules~\cite{Bruno};
local expansion, sign valid only for a positive triple
product~\cite{LiGarg} &
\raggedright $\det M<0$, Result~\ref{res:chern} \tabularnewline
\raggedright Sum rules & \raggedright topological~\cite{Bruno} &
\raggedright by counting, given Result~\ref{res:chern};
Sec.~\ref{sec:sumrules}
\end{tabular}
\end{ruledtabular}
\end{table}

% =====================================================================
\section{Rotated frames, the chain, and the lattice}
\label{sec:chain}
% =====================================================================

The derivation of the lattice in this section follows Ref.~\cite{KG} and
contains no new result; the section also fixes the notation used later and
gives a geometric reading of the two rotated axes. Throughout, $J=\tfrac12,1,\tfrac32,\dots$, $N=2J+1$, and
$0<k_2<k_1$, so that $c,s$ in \eqref{eq:Zpm} satisfy $c^2+s^2=1$ and
$0<c,s<1$. The uniaxial limits $k_2\to k_1$ ($s\to0$) and $k_2\to0$ ($c\to0$) are
singular: the axes of the two rotated frames introduced below become parallel
or antiparallel, so that the two chains coincide up to the ordering of the
sites, and the lattice contracts onto the $h_z$ or the $h_x$ axis,
respectively. Results~\ref{res:cuts}--\ref{res:chern} are not claimed at these
limits; Sec.~\ref{sec:chern} discusses the limit $k_2\to k_1$ for $J=1$. For $J=\tfrac12$ the
anisotropy is a multiple of the identity and the model is a free spin in a
field; the general formulas remain valid and give the single DP at $\bh=0$
(Appendix~\ref{app:chain}).

Using $S_x^2+S_y^2+S_z^2=J(J{+}1)\one$, Eq.~\eqref{eq:H} may also be written
$H=k_2J(J{+}1)\one+(k_1-k_2)S_x^2-k_2S_z^2-\bh\cdot\bS$, which is, up to the
constant, Eq.~(1) of Ref.~\cite{KG} and the standard molecular-magnet form
$-k_2J_z^2+(k_1-k_2)J_x^2-g\mu_B\mathbf J\cdot\mathbf H$ of
Ref.~\cite[Eq.~(81)]{GargAJP}, with $h_a=g\mu_BH_a$.

For $\varepsilon=\pm1$ let $X_\varepsilon,Y,Z_\varepsilon$ be the spin
components in the frame rotated about $y$ through $\varepsilon\theta$, where
$c=\cos\theta$ and $s=\sin\theta$,
\begin{equation}
  X_\varepsilon=cS_x-\varepsilon sS_z,\quad Y=S_y,\quad
  Z_\varepsilon=\varepsilon sS_x+cS_z ,
  \label{eq:rot}
\end{equation}
with $Z_{+1}=\Zp$ and $Z_{-1}=\Zm$, and rotate the field in the same way,
\begin{equation}
  u_\varepsilon=ch_x-\varepsilon sh_z,\quad v=h_y,\quad
  w_\varepsilon=\varepsilon sh_x+ch_z .
  \label{eq:rotfield}
\end{equation}
For this angle the $X_\varepsilon^2$ term cancels (Appendix~\ref{app:chain}),
leaving
\begin{align}
  H&=k_2J(J{+}1)\one+(k_1{-}2k_2)Z_\varepsilon^2
     +\varepsilon a\{X_\varepsilon,Z_\varepsilon\} \notag\\
   &\quad-u_\varepsilon X_\varepsilon-vY-w_\varepsilon Z_\varepsilon ,
  \label{eq:Hrot}
\end{align}
with $a=k_1cs=\sqrt{k_2(k_1-k_2)}>0$. Every term is diagonal or
nearest-neighbor in the $Z_\varepsilon$ eigenbasis, so the chain picture is
exact.

The two axes $\hat n_\pm=(\pm s,0,c)$ of $Z_\pm$ have a simple geometric
meaning. Restricted to the plane perpendicular to $\hat n_\pm$, the quadratic
form $k_1x^2+k_2y^2$ is $k_2$ times the identity: $\hat n_\pm$ are the normals
of the two circular central sections of every ellipsoid obtained from
$k_1x^2+k_2y^2=1$ by adding a positive multiple of $x^2+y^2+z^2$ to the
left-hand side, which changes the anisotropy only by a multiple of
$\bS^2=J(J{+}1)\one$. They are the analog of the optic axes of a biaxial
crystal, and the $\pi$ rotation about the easy axis $z$ exchanges them. Because the quadratic form is isotropic in the plane
perpendicular to each axis, the anisotropy has no transverse quadrupole
$X_\varepsilon^2-Y^2$ about that axis and no matrix elements with
$\Delta r=\pm2$.
The only off-diagonal anisotropy term left is
$\varepsilon a\{X_\varepsilon,Z_\varepsilon\}$, which on the bond
$r\leftrightarrow r+1$ has the same matrix element as a Zeeman term
$-h^{\rm eff}_rX_\varepsilon$ of the form in \eqref{eq:H}, with effective
transverse field $h^{\rm eff}_r=-\varepsilon a(2r+1)$. Classically, the
anisotropy energy of a unit vector $\hat e$ is
$k_2-k_1(\hat e\cdot\hat n_+)(\hat e\cdot\hat n_-)$, whose operator version is
the first line of Eq.~\eqref{eq:heunform} below.

\begin{figure}[tbp]
\centering
\includegraphics[width=0.94\columnwidth]{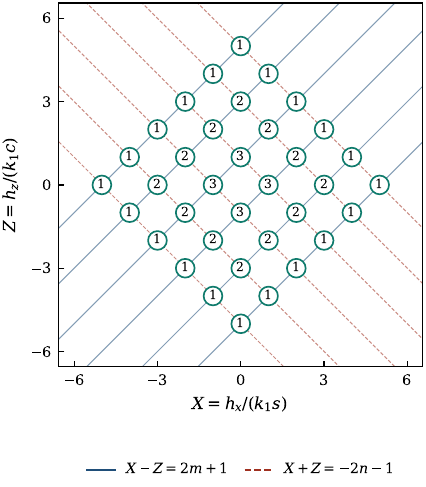}
\caption{The DP lattice for $J=3$ in the scaled coordinates
$X=h_x/(k_1s)$ and $Z=h_z/(k_1c)$. Solid lines: the $2J$ lines $X-Z=2m+1$,
$m=-J,\dots,J-1$, on which the $\Zp$ chain is cut at the bond
$m\leftrightarrow m+1$, so that $P_m$ commutes with $H$. Dashed lines: the
$2J$ lines $X+Z=-2n-1$ on which the $\Zm$ chain is cut at
$n\leftrightarrow n+1$, so that $Q_n$ commutes with $H$. Only at the
intersections do both projectors commute with $H$, and only there do DPs
occur. The
integer at each point is the number $f(m,n)$ of distinct doublets there,
Eq.~\eqref{eq:f}; it is constant on concentric rhombi.}
\label{fig:lattice}
\end{figure}

Let $\ket{r;\varepsilon}=e^{-\ii\varepsilon\theta S_y}\ket r$,
$r=-J,\dots,J$, where on the right-hand side $\ket r$ is the $S_z$ eigenstate
in the standard (Condon--Shortley) phase convention,
$\me{r{+}1}{S_+}{r}=\ell_r>0$ with $\ell_r$ given below. Then
$\ket{r;\varepsilon}$ is the eigenvector of $Z_\varepsilon$ with eigenvalue
$r$; this phase convention is used throughout, and off-diagonal matrix
elements depend on it. Below, $\ket r$ stands for $\ket{r;\varepsilon}$ when
the frame is clear. The hoppings are
\begin{equation}
  t^{(\varepsilon)}_r=\me{r{+}1;\varepsilon}{H}{r;\varepsilon}
   =\frac{\ell_r}{2}\bigl[\varepsilon a(2r+1)-u_\varepsilon+\ii v\bigr] ,
  \label{eq:hop}
\end{equation}
with $\ell_r=\sqrt{(J-r)(J+r+1)}>0$ for $-J\le r\le J-1$. A bond is cut when
the applied transverse field $(u_\varepsilon,v)$ cancels this effective field,
$u_\varepsilon+h^{\rm eff}_r=0$ and $v=0$, that is,
$u_\varepsilon=\varepsilon a(2r+1)$. Since $h^{\rm eff}_r$ is linear in $r$
with nonzero slope, each bond is cut at its own value of $u_\varepsilon$.
Requiring a
vanishing hopping in both frames at once gives $h_y=0$ together with
$ch_x-sh_z=a(2m+1)$ and $ch_x+sh_z=-a(2n+1)$, that is,
\begin{equation}
\begin{aligned}
  h^{(m,n)}_x&=k_1s\,(m-n),\\
  h^{(m,n)}_z&=-k_1c\,(m+n+1),\\
  h^{(m,n)}_y&=0 ,
\end{aligned}
\label{eq:lattice}
\end{equation}
with
\begin{equation}
  m,n\in\{-J,-J+1,\dots,J-1\} .
  \label{eq:mnrange}
\end{equation}
In the scaled field coordinates $X=h_x/(k_1s)$ and $Z=h_z/(k_1c)$, which are
numbers and not the spin components $X_\varepsilon$, $Z_\varepsilon$ of
Eq.~\eqref{eq:rot}, the point is
$(X,Z)=(m-n,-m-n-1)$, and the points form a centered rectangular lattice
(Fig.~\ref{fig:lattice}) with spacings $\Delta h_x=2k_1s$ and $\Delta
h_z=2k_1c$. We write $H_{mn}=H(\bh^{(m,n)})$ and record the ranks of the two
cut projectors \eqref{eq:PQ},
\begin{equation}
  p=J+m+1,\qquad q=J+n+1,\qquad 1\le p,q\le N-1 .
  \label{eq:pq}
\end{equation}
Equation~\eqref{eq:lattice} is a necessary condition on the field
(Appendix~\ref{app:chain}); by itself it does not say how many degeneracies
occur at a lattice point, nor that any occur.

% =====================================================================
\section{The two spectral cuts}
\label{sec:cuts}
% =====================================================================

Three kinds of operators carry the argument. All of them are fixed by the labels
$(m,n)$, and none requires the spectrum of $H$. The cut projectors $P_m$ and
$Q_n$ of Eq.~\eqref{eq:PQ} project onto the eigenspaces of $\Zp$ with eigenvalues $\le m$ and of $\Zm$
with eigenvalues $\le n$; they commute with $H_{mn}$ (Result~\ref{res:cuts}), and
$2P_m-1$, $2Q_n-1$ are the corresponding $\mathbb Z_2$ symmetries. The pairing
operator $B_{mn}=P_mQ_n(1-P_m)$, Eq.~\eqref{eq:B}, maps the partner of a
doublet in the upper segment of the $\Zp$ chain onto a nonzero multiple of its
partner in the lower segment, and its rank counts the doublets (Result~\ref{res:mult}). The
pseudospin operators $\tau$ of Eq.~\eqref{eq:tau} are built from $P_m$ and
$B_{mn}$ and act on the span of the doublets.

Since $\Zp$ and $\Zm$ have finite and simple spectra, the projectors
\eqref{eq:PQ} are explicit polynomials of degree $N-1$ in the corresponding
spin component,
\begin{equation}
  P_m=\sum_{r=-J}^{m}\prod_{\substack{t=-J\\t\neq r}}^{J}\frac{\Zp-t\one}{r-t} ,
  \label{eq:Ppoly}
\end{equation}
and similarly for $Q_n$.

\begin{result}[Two cuts, each commuting with $H$]
\label{res:cuts}
At the lattice point \eqref{eq:lattice},
$[H_{mn},P_m]=[H_{mn},Q_n]=0$. In the $\Zp$ basis exactly one hopping of
$H_{mn}$ vanishes, namely across the bond $m\leftrightarrow m+1$, and all
others are nonzero. The dual statement holds in the $\Zm$ basis with the cut at
$n$.
\end{result}

\noindent Result~\ref{res:cuts} follows from the form of $H_{mn}$ that treats
the two rotated axes symmetrically,
\begin{align}
  H_{mn}&=k_2J(J{+}1)\one-\frac{k_1}{2}\{\Zp,\Zm\} \notag\\
        &\quad+k_1\Bigl(n+\tfrac12\Bigr)\Zp
              +k_1\Bigl(m+\tfrac12\Bigr)\Zm ,
  \label{eq:heunform}
\end{align}
obtained by inserting \eqref{eq:lattice} into \eqref{eq:H} with
$S_x=(\Zp-\Zm)/(2s)$, $S_z=(\Zp+\Zm)/(2c)$ and
$\{\Zp,\Zm\}=2c^2S_z^2-2s^2S_x^2$. In the $\Zp$ basis,
$\Zm=\cos2\theta\,\Zp-\sin2\theta\,X_+$ is itself an irreducible chain, with
hoppings $(\Zm)_{r+1,r}=-\sin2\theta\,\ell_r/2\neq0$ for $0<\theta<\pi/2$;
here and below $A_{r+1,r}$ denotes $\me{r{+}1}{A}{r}$. The off-diagonal
elements of \eqref{eq:heunform} come from $\Zm$, either directly or through the
anticommutator, $\{\Zp,\Zm\}_{r+1,r}=(2r+1)(\Zm)_{r+1,r}$. Collecting them,
the hopping of $H_{mn}$ across $r\leftrightarrow r+1$ is
$[-\tfrac{k_1}{2}(2r+1)+k_1(m+\tfrac12)](\Zm)_{r+1,r}=k_1(m-r)(\Zm)_{r+1,r}$,
which vanishes only at $r=m$. A tridiagonal matrix whose hopping across
$m\leftrightarrow m+1$ vanishes is block diagonal with respect to
$\HS=P_m\HS\oplus(1-P_m)\HS$, so $[H_{mn},P_m]=0$. Exchanging
$\Zp\leftrightarrow\Zm$ and $m\leftrightarrow n$ gives the statement for
$Q_n$.

Each cut by itself is the two-block decomposition of Ref.~\cite{KG}. That work
used both rotated frames to locate the lattice: the vanishing hopping in one
frame gives its Eq.~(9), in the mirrored frame its Eq.~(10), and the blocks
of the two frames have sizes $q,N-q$ and $p,N-p$ under the index dictionary
given in the Supplemental Material~\cite{supp}. Reference~\cite{KG} did not
consider the two decompositions as two operators acting together, and its
counting argument uses the blocks of one frame only. The relation between the
two cuts is the subject of Results~\ref{res:transversal} and~\ref{res:mult}:
the two projectors do not commute, and the rank of the pairing operator built
from them gives the multiplicities.

The pair $(\Zp,\Zm)$ is a Krawtchouk--Leonard pair: each member has a simple,
equally spaced spectrum and is irreducibly tridiagonal in an eigenbasis of the
other~\cite{NomuraTerwilliger}. Equation~\eqref{eq:heunform} is the algebraic
Heun operator of that pair, bilinear in the two members, with the coefficient
of $\Zp$ locked to $k_1(n+\tfrac12)$ and that of $\Zm$ to $k_1(m+\tfrac12)$.
An operator of this kind, with coefficients fixed so that it commutes with one
spectral cutoff of each member of a bispectral pair, and hence with
the time--band limiting operator formed from the two cutoffs, is the commuting
tridiagonal matrix that Perline~\cite{Perline} constructed for finite
time--band limiting; in
the present form it is Eq.~(5.34) of Gr\"unbaum, Vinet and Zhedanov with their
conditions (5.31)--(5.33)~\cite{GVZ}, see also Refs.~\cite{BCV,ReviewFF}. In
the $\mathfrak{su}(2)$ case the operator without the locking, Eq.~(3.13) of
Ref.~\cite{CVZ}, has been identified there with the Hamiltonian of a quantum
analog of the Zhukovski--Volterra gyrostat; for the pair $\Zp/c$, $\Zm/c$ it is
$H(\bh)$ of Eq.~\eqref{eq:H} at arbitrary field. The operator itself is
therefore known. New here is that the locking conditions hold exactly at the
points of the Ke\c{c}ecio\u{g}lu--Garg lattice, with the two lattice labels as
the two cutoffs. A Hermitian operator affine in each member of the pair reads
$\lambda\{\Zp,\Zm\}+\lambda'\,\ii[\Zp,\Zm]+\lambda_+\Zp+\lambda_-\Zm+\lambda_0\one$
with real coefficients. Since $\ii[\Zp,\Zm]=2csS_y$ is a field along $y$,
commutation with both cuts forces $\lambda'=0$, which is the condition $h_y=0$,
and $\lambda_+=-\lambda(2n+1)$, $\lambda_-=-\lambda(2m+1)$~\cite{supp};
$\lambda=-k_1/2$ and $\lambda_0=k_2J(J+1)$ give \eqref{eq:heunform}. In time--band limiting only the
restriction of the operator to a single block is used, so degeneracies between
the two blocks, which are the subject here, do not arise there. For a field along a principal axis, \eqref{eq:H} is also, up to a constant,
an anisotropic Lipkin--Meshkov--Glick Hamiltonian~\cite{LMG}.

\emph{In what sense the cuts are a hidden symmetry.} Both projectors are known
before any eigenvalue problem is solved: by Eq.~\eqref{eq:Ppoly} they are
polynomials in $\Zp$ and $\Zm$, which do not depend on the field. $P_m$
commutes with $H(\bh)$ exactly on its cut line, $h_y=0$ and $X-Z=2m+1$, and
$Q_n$ exactly on $h_y=0$ and $X+Z=-2n-1$ (Fig.~\ref{fig:lattice}), so that both
commute with $H$ only at $\bh^{(m,n)}$. They are therefore
not symmetries of the family $H(\bh)$, and a search for a symmetry of the whole
family does not find them, but $2P_m-1$ and $2Q_n-1$ are unitary involutions
commuting with $H_{mn}$. Symmetries attached to particular parameter values
also occur elsewhere: in the asymmetric quantum Rabi model, whose levels cross when the bias is a
multiple of half the mode frequency, the crossings are traced to operators that
commute with the Hamiltonian at these bias values, constructed in
Ref.~\cite{Rabi} for the first several of them.
Reference~\cite{GargAJP} draws the same distinction in general, noting in
its discussion of codimensions that the argument ``does not, however, preclude
the existence of a symmetry at the degeneracy point itself''.

At a given field a single cut is a block decomposition and does not by itself
force a degeneracy; along its cut line, where $2P_m-1$ is a $\mathbb Z_2$
symmetry, the crossings that the ordering of levels requires
(Appendix~\ref{app:levels}) all sit at lattice points. There the pair implies
degeneracy because $[P_m,Q_n]\neq0$ (Result~\ref{res:transversal}), by the
following elementary fact. If $H$ has a simple spectrum,
$H=\sum_kE_k\ket k\bra k$ with distinct $E_k$, then $[A,H]=0$ gives
$(E_k-E_l)\me kAl=0$, so every operator commuting with $H$ is diagonal in the
eigenbasis of $H$, and any two such operators commute. Two operators that
commute with $H$ but not with each other therefore force a degenerate
eigenvalue. Quantitatively, $[P_m,Q_n]$ commutes with $H_{mn}$ and vanishes on
every one-dimensional eigenspace, on which $P_m$ and $Q_n$ act as numbers, so
its range lies in the span of the degenerate eigenspaces; their total
dimension is at least $\rank[P_m,Q_n]$. The two noncommuting involutions
generate a non-Abelian group, and the algebra they generate contains a
pseudospin SU(2) on the degenerate subspace that commutes with $H_{mn}$,
Eq.~\eqref{eq:factorization}.

% =====================================================================
\section{The number of degeneracies}
\label{sec:counting}
% =====================================================================

\subsection{Relative position of the two cuts}

Because both cuts commute with the Hamiltonian, so does
\begin{equation}
  B_{mn}=P_mQ_n(1-P_m) ,
  \label{eq:B}
\end{equation}
which projects a state onto the upper segment of the $\Zp$ chain, applies
$Q_n$, and projects the result onto the lower segment. It vanishes identically
when the two cuts commute, and its rank measures how far they are from doing
so. Inserting $P_m+(1-P_m)$ on either side of $Q_n$ gives
\begin{equation}
  [P_m,Q_n]=B_{mn}-B_{mn}^\dagger .
  \label{eq:commB}
\end{equation}
The two terms map between the two orthogonal segments in opposite directions,
so the commutator has twice the rank of $B_{mn}$.

\begin{result}[Exact transversality]
\label{res:transversal}
For every pair of labels \eqref{eq:mnrange} the two cuts are in general
position, that is, the four common eigenspaces of $P_m$ and $Q_n$ have the
smallest dimensions compatible with the ranks $p$ and $q$, and
\begin{equation}
  \rank B_{mn}=f(m,n),
  \qquad
  \rank[P_m,Q_n]=2f(m,n),
  \label{eq:rank}
\end{equation}
where
\begin{align}
  f(m,n)&=\min\{p,\,q,\,N-p,\,N-q\} \notag\\
        &=\tfrac12\bigl(N-|m-n|-|m+n+1|\bigr)\ \ge\ 1 .
  \label{eq:f}
\end{align}
\end{result}

\begin{figure}[tbp]
\centering
\includegraphics[width=0.94\columnwidth]{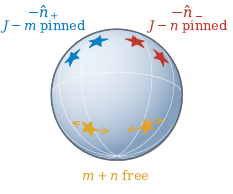}
\caption{Counting stars. A spin-$J$ state is represented by $2J$ points on the
sphere. States in the range of $P_m$ carry at least $J-m$ stars at
$-\hat n_+$, those in the range of $Q_n$ at least $J-n$ at $-\hat n_-$,
and the two directions are distinct; the stars of each cluster coincide and
are drawn apart only for visibility. A state lying in both must carry both
clusters. For $m+n\ge0$ this leaves $2J-(J-m)-(J-n)=m+n$ stars free, and such
states form a space of dimension $m+n+1$; for $m+n<0$ there is no such
state.}
\label{fig:stars}
\end{figure}

The proof (Appendix~\ref{app:stars}) counts Majorana stars. In the stellar
representation~\cite{Majorana} a spin-$J$ state is represented, up to normalization and phase, by $2J$
points on the unit sphere, counted with multiplicity, and the range of a cut projector is the set
of states carrying at least a prescribed number of stars at a prescribed point:
$\Ran P_m$ consists of the states with at least $J-m$ stars at $-\hat n_+$, and
$\Ran Q_n$ of those with at least $J-n$ stars at $-\hat n_-$. The two
directions are distinct, so a state in both ranges must carry two separate
clusters, and only $2J$ stars are available (Fig.~\ref{fig:stars}). Demands
for $\xi$ and $\eta$ stars at two different points can be met simultaneously if
and only if $\xi+\eta\le2J$, and the states meeting them form a space of
dimension $2J-\xi-\eta+1$. This fixes the dimensions of the four common eigenspaces of $P_m$
and $Q_n$ at the smallest values compatible with $p$ and $q$, from which
\eqref{eq:rank} follows. The four dimensions also follow from the corner
minors of the Wigner rotation matrix, which in addition bound the nonzero
singular values of $B_{mn}$ from below~\cite{supp}, and the multiplicity
$f(m,n)$ follows independently from the ordering of levels along a cut line
(Appendix~\ref{app:levels}).

\subsection{Exact multiplicities}

The number $f(m,n)$ in \eqref{eq:f} is the multiplicity found by
Ke\c{c}ecio\u{g}lu and Garg~\cite[Eq.~(17)]{KG}.

\begin{result}[Exact multiplicities]
\label{res:mult}
$H_{mn}$ has exactly $f(m,n)$ distinct twofold-degenerate eigenvalues. Every
other eigenvalue is simple, and no eigenvalue has multiplicity greater than
two. The degeneracies of \eqref{eq:H} therefore occur exactly at the lattice
points \eqref{eq:lattice}, and each lattice point carries $f(m,n)$ DPs at
distinct energies, every one of which is a linear conical intersection by
Result~\ref{res:chern}.
\end{result}

The proof has three steps. Let $\HS_{ij}$, $i,j\in\{0,1\}$, be the common
eigenspace on which $P_m$ takes the value $i$ and $Q_n$ the value $j$, with the
dimensions $d_{ij}$ of Eq.~\eqref{eq:dij}.

(i) \emph{No eigenvalue is more than twofold degenerate.} Each of the two
segments of the severed $\Zp$ chain is an irreducible chain
(Result~\ref{res:cuts}) and has a simple spectrum
(Appendix~\ref{app:chain}). Since $P_m$ commutes with $H_{mn}$, every
eigenspace $\mathcal E$ of $H_{mn}$ is the direct sum of its parts in the two
segments, each at most one-dimensional. Hence $\dim\mathcal E\le2$, and a
doublet has an orthonormal basis $\ket L\in P_m\HS$, $\ket R\in(1-P_m)\HS$.

(ii) \emph{There are at least $f(m,n)$ doublets.} $B_{mn}$ commutes with
$H_{mn}$ and therefore maps every eigenspace into itself, so $\rank B_{mn}$ is
the sum of the ranks of its restrictions. On a one-dimensional eigenspace the
restriction vanishes, because $B_{mn}^2=0$ [from $(1-P_m)P_m=0$]. On a doublet
it is $\me L{Q_n}R\,\ket L\bra R$, of rank one or zero. With
Result~\ref{res:transversal}, the number of doublets is at least
$\rank B_{mn}=f(m,n)$.

(iii) \emph{Every doublet contributes one unit of rank.} Suppose that
$\me L{Q_n}R=0$ for some doublet. $Q_n$ maps the doublet into itself, and its
restriction is a Hermitian $2\times2$ matrix in the basis $(\ket L,\ket R)$
whose off-diagonal entries are $\me L{Q_n}R$ and its complex conjugate. The
matrix is therefore diagonal, and since $Q_n^2=Q_n$ its diagonal entries are
$0$ or $1$, so that $\ket L\in\HS_{1j}$ and $\ket R\in\HS_{0j'}$ for some
$j,j'$. If $j=j'$, then $\ket L$ and $\ket R$ are two orthogonal eigenvectors
with the same energy in one segment of the severed $\Zm$ chain, whose spectrum is simple (Result~\ref{res:cuts} and
Appendix~\ref{app:chain}): a contradiction. If $(j,j')=(1,0)$, then
$\HS_{11}$ and $\HS_{00}$ are both nonzero, which by \eqref{eq:dij} requires
$p+q>N$ and $p+q<N$; if $(j,j')=(0,1)$, then $\HS_{10}$ and $\HS_{01}$ are both
nonzero, which requires $p>q$ and $q>p$. Each case is impossible, so every
doublet contributes exactly one unit to $\rank B_{mn}$, and there are exactly
$f(m,n)$ doublets.

The same argument identifies the paired subspace $\Hpair$, the span of all
doublets, without diagonalizing $H_{mn}$. A simple eigenvalue has an
eigenvector in one of the $\HS_{ij}$, on which $P_m$ and $Q_n$ act as numbers.
There are $N-2f$ such eigenvectors, and $\sum_{ij}d_{ij}=N-2f$, so they span
$\bigoplus_{ij}\HS_{ij}$, and $\Hpair$ is its orthogonal complement. The range
of $[P_m,Q_n]$ lies in $\Hpair$ (Sec.~\ref{sec:cuts}) and has dimension $2f$,
so $\Hpair=\Ran[P_m,Q_n]$. The counting is done at the physical anisotropy and
replaces the deformation and continuity argument of Ref.~\cite{KG}; step (i),
which uses only one of the two cuts, is already implicit in that work.

The pairing operator also supplies the pseudospin. Let
\begin{equation}
  \tau_+=B_{mn}\bigl(B_{mn}^\dagger B_{mn}\bigr)^{-1/2},\qquad
  \tau_-=\tau_+^\dagger,
  \label{eq:tau}
\end{equation}
where the inverse square root is taken on the
range of $B_{mn}^\dagger B_{mn}$, which is $\Hpair\cap(1-P_m)\HS$, and $\tau_+$
is set to zero on its orthogonal complement. Then $\tau_z=[\tau_+,\tau_-]$
equals $2P_m-1$ on $\Hpair$ and vanishes on its orthogonal complement, and we
set $\tau_x=\tau_++\tau_-$ and $\tau_y=-\ii(\tau_+-\tau_-)$. In a basis of doublet partners
$\ket{L_\nu}$, $\ket{R_\nu}$, $\nu=1,\dots,f$, with phases chosen so that
$\me{L_\nu}{Q_n}{R_\nu}>0$, one has $\tau_+=\sum_\nu\ket{L_\nu}\bra{R_\nu}$
\cite{supp}. The $\tau$ obey the Pauli algebra on $\Hpair$, and being functions
of $P_m$ and $Q_n$ alone they commute with $H_{mn}$, so
\begin{equation}
  \Hpair\simeq\mathbb C^2\otimes\mathbb C^{f(m,n)},
  \qquad
  H_{mn}\big|_{\Hpair}=\one_2\otimes K_{mn} ,
  \label{eq:factorization}
\end{equation}
with $K_{mn}$ of simple spectrum. For $J\ge1$ the pseudospin is distinct from
the physical spin: $\tau_+$ is nilpotent of rank $f(m,n)\le N/2<2J$, whereas a
nonzero nilpotent combination of $\one,S_x,S_y,S_z$ is a raising operator about
some axis and has rank $2J$. For $J=\tfrac12$, where $\Hpair=\HS$, the $\tau_i$
are, up to signs, twice the spin components in the $\Zp$ frame. For every $J$
the $\tau_i$ commute with $H(\bh)$ only at $\bh^{(m,n)}$: otherwise they would
commute with $H(\bh)-H_{mn}=-(\bh-\bh^{(m,n)})\cdot\bS$, which would then leave
$\Hpair$ invariant and act on it as $\one_2\otimes K'$, so that $(\bh-\bh^{(m,n)})\cdot\bS$
would have a degenerate eigenvalue, whereas its spectrum,
$|\bh-\bh^{(m,n)}|\times\{-J,\dots,J\}$, is simple. The two basis states of the
pseudospin are the two segments of the severed $\Zp$ chain, the eigenvalues
$\pm1$ of $2P_m-1$.

For half-integer $J$ the point $m=n=-\tfrac12$ lies at $\bh=0$, with $p=q=N/2$
and $f=N/2$, so the entire spectrum consists of $N/2$ doublets and
$\Hpair=\HS$; Ref.~\cite{KG} already used this case as a consistency check.
These doublets are the Kramers doublets, which time reversal guarantees for
every time-reversal-invariant Hamiltonian at zero field. The two-cut pairing is
a different mechanism that also applies at this point: it is unitary rather
than antiunitary, it depends on the nearest-neighbor chain and hence on the
quadratic anisotropy, and for $J\ge\tfrac52$ it is lost when the fourth-order term
$-C(S_+^4+S_-^4)$ of Sec.~\ref{sec:beyond} is added, whereas the Kramers
degeneracy is not.

\subsection{Sum rules}
\label{sec:sumrules}

Summing \eqref{eq:f} over the $(2J)^2$ lattice points gives the total number of
degenerate level pairs,
\begin{equation}
  \sum_{m,n=-J}^{J-1}f(m,n)=\frac{2J(J+1)(2J+1)}{3} ,
  \label{eq:total}
\end{equation}
which exceeds the number $(2J)^2$ of lattice points by $\tfrac23J(J-1)(2J-1)$
for every $J\ge\tfrac32$; for $J=\tfrac12$ and $J=1$ the two are equal.
Counting the crossings on each cut line (Appendix~\ref{app:levels}) gives the
number of DPs joining the globally ordered levels $k$ and $k+1$,
\begin{equation}
  N_k=k(N-k),\qquad k=1,\dots,N-1 ,
  \label{eq:Nk}
\end{equation}
and the row identity
\begin{equation}
  \sum_{n=-J}^{J-1}f(m,n)=p\,(N-p) ,
  \label{eq:rowsum}
\end{equation}
which states that the $p(N-p)$ crossings between the two segments on the cut
line $m$ are distributed over the $2J$ lattice points of that line as $f(m,n)$
prescribes. Equations~\eqref{eq:total} and \eqref{eq:Nk} are not new. Once
every DP is known to have Bruno's diabolicity index $+1$, which is $\chi=-1$
in the convention of Sec.~\ref{sec:chern} (Result~\ref{res:chern}), they are
Bruno's sum rules for the summed and the level-resolved diabolicity
index~\cite[Eqs.~(2b), (2a)]{Bruno}, specialized to this model:
Ref.~\cite{Bruno} labels the levels by $\mu=J$ for the ground state down to
$\mu=-J$, so that the pair $k,k+1$ carries $\mu=J+1-k$ and the right-hand side
of Eq.~(2a) there is $(J+\mu)[J-(\mu-1)]=k(N-k)$. Bruno obtained them from the
topology of the Berry curvature; the derivation here counts level crossings
and uses no topological input, and Result~\ref{res:chern} gives the index of
each point separately rather than only their sums.

The case $k=1$ is the pair whose tunnel splitting is measured at $h_z=0$, and
\eqref{eq:Nk} gives $N_1=2J$ DPs between the two lowest levels.
Appendix~\ref{app:levels} shows that on the cut line $m$ the crossing of the
$i$-th level of the lower segment with the $j$-th level of the upper segment
sits at the lattice point whose second index is
\begin{equation}
  n=i-j-m-1,\quad\text{i.e.}\quad i-j=m+n+1=p+q-N .
  \label{eq:whichpoint}
\end{equation}
A crossing $(i,j)$ has $(i-1)+(j-1)$ levels below it and so joins the global
levels $i+j-1$ and $i+j$ (Appendix~\ref{app:levels}); a DP of the two lowest
levels on the cut line is therefore the crossing $i=j=1$, and it lies at
$n=-m-1$, where \eqref{eq:lattice} gives
\begin{equation}
  h_z=0,\qquad h_x=k_1s\,(2m+1),\qquad m=-J,\dots,J-1 .
  \label{eq:groundDP}
\end{equation}
All $2J$ of them lie on the hard axis, at $h_x/(k_1s)=\pm1,\pm3,
\dots,\pm(2J-1)$ for integer $J$ and at $h_x/(k_1s)=0,\pm2,\dots,\pm(2J-1)$ for
half-integer $J$, where the point $h_x=0$ is the Kramers point. These
are the quenching fields obtained by
Garg~\cite{Garg1993,GargPRL1999,Garg2000,GargI}; the
count adds that no DP between the two lowest levels lies anywhere else. For
$J=10$ the formula gives twenty points, ten of them at positive $h_x$.
Four of the ten, shifted along the axis by anisotropy of fourth order, were
resolved in the tunnel-splitting oscillations of
Fe$_8$~\cite{WernsdorferSessoli}; the same term replaces the remaining six by
three mirror pairs off the hard axis~\cite{Bruno,LiGarg}
(Sec.~\ref{sec:beyond}).

% =====================================================================
\section{Orientation of the cones}
\label{sec:chern}
% =====================================================================

A DP is a Berry-curvature monopole. Its charge has unit modulus whenever the
intersection is a nondegenerate linear cone, and its sign, the orientation of
the cone, is a separate question. Bruno's sum rules~\cite{Bruno}, combined with
the Ke\c{c}ecio\u{g}lu--Garg count and with the assumption that every index has
modulus one, force a common value at every point; the linear-cone property
behind that modulus had not been proven. Li and Garg~\cite{LiGarg} expanded
around a DP, but the sign they obtain holds only when a certain triple product
is positive, which their flux integral takes for granted (see below);
Result~\ref{res:chern} gives that triple product exactly, in factorized form,
for every doublet.

Let $\mathcal E$ be a doublet, $(\ket{+},\ket{-})$ an orthonormal basis of it,
and $(h_1,h_2,h_3)$ an oriented orthonormal field frame with conjugate
operators $T_a=-\partial H/\partial h_a$, which for the Zeeman coupling of
\eqref{eq:H} are the spin components along the frame axes. The splitting
Jacobian $M$ is
\begin{align}
  M_{1a}&=-\operatorname{Re}\me{+}{T_a}{-}, \notag\\
  M_{2a}&=\operatorname{Im}\me{+}{T_a}{-}, \label{eq:Mdef}\\
  M_{3a}&=-\tfrac12\bigl(\me{+}{T_a}{+}-\me{-}{T_a}{-}\bigr), \notag
\end{align}
so that, to first order in $\delta\bh$, the doublet Hamiltonian is $(M\delta\bh)\cdot\bm\sigma$
up to a multiple of the identity, and the gap is
$2|M\delta\bh|+O(|\delta\bh|^2)$. Thus $\det M\neq0$ is the condition for an
isolated linear cone. Write $\mathcal A=\ii\langle\psi|\nabla\psi\rangle$ for the
Berry connection of a level $\ket\psi$ and $\mathcal F=\nabla\times\mathcal A$ for its
curvature, and set
\begin{equation}
  \chi=\frac{1}{2\pi}\oiint\mathcal F\cdot d\bm\Sigma
      =\sgn\det M
  \label{eq:convention}
\end{equation}
for the Chern charge of the lower level, the integral running over a small
sphere about the DP with outward normal. The second equality, valid whenever
$\det M\neq0$, holds because the map
$\delta\bh\mapsto\mathbf d/|\mathbf d|$, with $\mathbf d=M\delta\bh$ the Pauli
vector, takes the small sphere onto the unit sphere with degree $\sgn\det M$,
and to leading order the lower eigenvector depends on $\delta\bh$ only through
$\mathbf d/|\mathbf d|$, so that its curvature is the pullback of the
curvature on the unit sphere~\cite{Berry1984,GargAJP}. For $M=\one$, that is,
$\mathbf d=\delta\bh$, the lower level of $H=\mathbf d\cdot\bm\sigma$ has
$\mathcal F=\mathbf d/(2|\mathbf d|^3)$ and $\chi=+1$, the standard case of
Ref.~\cite{Berry1984}. A negative $\det M$
means that $\delta\bh\mapsto M\delta\bh$ reverses orientation, and then the
lower level has $\chi=-1$. The sign is well defined once the field frame is
right-handed and the spin components are the generators in that same frame,
since a change of orthonormal basis in $\mathcal E$ acts on the Pauli vector
through SO(3) and a rotation of the field frame is likewise unimodular.

References~\cite{Bruno} and~\cite{LiGarg} define the charge of a level
$\ket\psi$ through the curvature
$\bm{\mathcal B}=-\operatorname{Im}\sum_{\psi'\neq\psi}\me{\psi}{\nabla H}{\psi'}\times
\me{\psi'}{\nabla H}{\psi}/(E_\psi-E_{\psi'})^2$ as
$\mathcal Q=-(2\pi)^{-1}\oiint\bm{\mathcal B}\cdot d\bm\Sigma$, which is
$\mathcal Q=-\chi$. For a two-level DP Bruno's diabolicity index is the
$\mathcal Q$ of the lower level, so his index unity is $\chi=-1$, in agreement
with Result~\ref{res:chern} below. Li and Garg quote \mbox{$\mathcal Q=-1$} for the
lower level. Their face normals are oriented correctly, but the surface element
of their Eq.~(19) carries the sign of the triple product $\Omega$ of their three
expansion vectors and points outward only for $\Omega>0$. In their notation
$\Omega=8\det M/J^3$, which is negative at every DP of this model, and with
$|\Omega|$ their computation gives \mbox{$\mathcal Q=+1$}, again $\chi=-1$. Bruno's sum
rule requires this value: the ground level is the lower level at all $N_1=2J$
of its DPs, and his Eq.~(2a) at $\mu=J$ gives $\sum_i\mathcal Q_i=2J$. Below we
quote $\chi$ throughout.

At $\bh^{(m,n)}$ the matrix of $H_{mn}$ in the $\Zp$ basis is real symmetric,
so its eigenvectors may be chosen real, and by step~(i) of the proof of Result~\ref{res:mult} the two
doublet partners may be taken to live one on each side of the cut,
$\ket{L}\in\HS_L=\mathrm{span}\{\ket r\}_{r\le m}$ and
$\ket{R}\in\HS_R=\mathrm{span}\{\ket r\}_{r\ge m+1}$. The amplitudes of these
states at the two sites adjoining the cut, $\alpha=\ip{m}{L}$ and
$\beta=\ip{m+1}{R}$, are nonzero, because an eigenstate of an irreducible chain
cannot vanish at an end site (Appendix~\ref{app:chain}). The two states also
occupy disjoint, ordered ranges of the rotated quantum number:
$\langle\Zp\rangle_L=\sum_{r\le m}r\,|\ip{r}{L}|^2\le m$ and
$\langle\Zp\rangle_R\ge m+1$, so that
\begin{equation}
  \langle \Zp\rangle_L\le m<m+1\le\langle \Zp\rangle_R,
  \qquad
  \langle \Zp\rangle_R-\langle \Zp\rangle_L\ge1 .
  \label{eq:zgap}
\end{equation}

\begin{result}[Uniform orientation]
\label{res:chern}
For every doublet at every lattice point,
\begin{equation}
  \det M=-\frac{\ell_m^2\,\alpha^2\beta^2}{8}
          \bigl(\langle \Zp\rangle_R-\langle \Zp\rangle_L\bigr)\;<\;0 .
  \label{eq:detM}
\end{equation}
Every degeneracy of \eqref{eq:H} is therefore an isolated linear cone, and its
lower level carries Chern charge $\chi=-1$.
\end{result}

\noindent We use the doublet basis $(\ket L,\ket R)$ and the field frame
$(u_{+1},v,w_{+1})$ of \eqref{eq:rotfield}, whose subscript we drop, with
conjugate operators $(X_+,Y,\Zp)$. The rotation
$(h_x,h_y,h_z)\mapsto(u,v,w)$ has determinant $c^2+s^2=+1$. The operators $X_+$
and $Y$ are nearest-neighbor in the $\Zp$ basis, so their only matrix elements
between $\HS_L$ and $\HS_R$ are across the bond $m\leftrightarrow m+1$, while
$\Zp$ is diagonal. With $\gamma=\tfrac12\ell_m\alpha\beta\neq0$ this gives
$\me{L}{X_+}{R}=\gamma$, $\me{L}{Y}{R}=\ii\gamma$ and $\me{L}{\Zp}{R}=0$, and
$\me{L}{Y}{L}=\me{R}{Y}{R}=0$ because the states are real and $Y$ is purely
imaginary and antisymmetric in the $\Zp$ basis. Definition \eqref{eq:Mdef}
then gives
\begin{equation}
  M_{uvw}=
  \begin{pmatrix}
    -\gamma & 0 & 0\\
    0 & \gamma & 0\\
    \tfrac12(\langle X_+\rangle_R-\langle X_+\rangle_L) & 0 &
    \tfrac12(\langle \Zp\rangle_R-\langle \Zp\rangle_L)
  \end{pmatrix} ,
\end{equation}
which is lower triangular; its determinant
$-\tfrac12\gamma^2(\langle\Zp\rangle_R-\langle\Zp\rangle_L)$ is
\eqref{eq:detM}. Both $\gamma^2$ and
$\langle\Zp\rangle_R-\langle\Zp\rangle_L$ are positive, by $\gamma\neq0$ and
\eqref{eq:zgap}. The matrix $M$ is the first-order splitting matrix of
degenerate perturbation theory, and \eqref{eq:detM} evaluates its determinant
exactly. At a lattice point with $f(m,n)>1$ the formula applies separately to
each of the $f$ doublets, which lie at distinct energies, so every one of them
is a cone.

The determinant \eqref{eq:detM} is strictly negative, but for ${J\ge1}$, $|\det M|$ has no anisotropy-independent positive lower bound. As $s\to0$,
$H_{mn}/k_1$ tends to $J(J{+}1)\one-S_z^2+(m+n+1)S_z$, which is diagonal in the limiting $\Zp$ basis with entries $J(J{+}1)+D_r$, where
$D_r=-r^2+(m+n+1)r$. A limit of $\ket L$ or
$\ket R$ is an eigenvector of this matrix for the limiting doublet energy, so
$\alpha$ and $\beta$ can both stay away from zero only if $D_m=D_{m+1}$, that
is, $m=n$. At every lattice point with $m\neq n$ (such points exist for $J\ge1$),
$\alpha\beta\to0$ and hence $\det M\to0$; as $c\to0$, where
$H_{mn}/k_1\to\Zp^2+(n-m)\Zp$, the same holds at every point with
$m+n+1\neq0$. For $J=1$, for instance, $\det M=-s^2/(1+s^2)^2$ at
$(m,n)=(0,-1)$; for $J=\tfrac12$, $\det M=-\tfrac18$ at every anisotropy. The determinant is
dimensionless and depends only on $J$, on $k_2/k_1$ and on the doublet, since
$H_{mn}/k_1$ depends on the anisotropy only through $k_2/k_1$. A small
$|\det M|$ means that the splitting grows slowly away from the DP in at least
one field direction; it does not indicate a nonconical degeneracy or a merging
of DPs. Section~\ref{sec:numerics} gives its size.

In the singular limit $k_2\to k_1$ at $J=1$ the two hard-axis points
$(m,n)=(0,-1)$ and $(-1,0)$, which are the only DPs joining the two lowest
levels ($N_1=2$), converge on $\bh=0$ from $h_x=\pm k_1s$. Once $s<\tfrac12$, both lie inside the sphere $|\bh|=k_1/2$, on which the two
lowest levels then stay nondegenerate up to and including $k_2=k_1$, where they meet only at $\bh=0$. The flux of the lower level
through that sphere is therefore constant, equal to $-2$ by
Result~\ref{res:chern}, and at $k_2=k_1$ it is the flux of the
Chern-number-two crossing of Ref.~\cite{GargAJP}, Sec.~III, which therefore has
$\chi=-2$ in the present convention. Cones of opposite orientation would have
given zero.

% =====================================================================
\section{Numerical verification}
\label{sec:numerics}
% =====================================================================

The proofs of Results~\ref{res:cuts}--\ref{res:chern} use no numerical input.
This section reports an independent numerical check of the commutation
relations, ranks, multiplicities, and the sign and factorized form of $\det M$ stated
in Results~\ref{res:cuts}--\ref{res:chern}, and quantifies $|\det M|$, for
which we have no closed-form lower bound. We
diagonalized
\eqref{eq:H} at every lattice point for $J=1,\tfrac32,\dots,4$ and
$k_2/k_1\in\{0.05,0.3,0.9\}$, built $P_m$ and $Q_n$ from the spectral
decompositions of $Z_\pm$, and evaluated \eqref{eq:Mdef} on each degenerate
pair in the original field frame $(h_x,h_y,h_z)$ rather than the rotated one,
so that the frame independence of $\sgn\det M$ is tested as well. Ranks are
taken with a singular-value cutoff of $10^{-13}$, and the run aborts unless every retained singular value exceeds the cutoff,
and the cutoff every discarded one, by at least a factor of ten; over the sweep the smallest retained
singular value is $2.0\times10^{-7}$ and the largest discarded one
$2.2\times10^{-15}$. Any cutoff between these two values, a window of eight
decades, gives the same ranks. The ranks themselves are proven analytically
(Result~\ref{res:transversal}), and the numerical classification rests on
this gap; the closed lower bound on the singular values given in the
Supplemental Material~\cite{supp} is far from sharp near the uniaxial limits.
Table~\ref{tab:summary} collects the global checks; all thresholds and the
complete point-by-point data for $J=2$ at $k_2/k_1=0.3$ are in the
Supplemental Material~\cite{supp}.

\begin{table}[!t]
\caption{Global checks. ``pts'' is the number of lattice points $(2J)^2$, that
is, of distinct fields at which DPs occur, and ``doublets'' the total number of
degenerate pairs found, equal at every anisotropy both to $\sum_{m,n}f(m,n)$
and to the closed form $\tfrac23J(J{+}1)(2J{+}1)$. The last three columns give
the largest (least negative) $\det M$ over the whole lattice at the indicated
anisotropy. All rank, multiplicity and sign checks passed in every row.}
\label{tab:summary}
\begin{ruledtabular}
\begin{tabular}{rrrrrrr}
 & & & & \multicolumn{3}{c}{$\max\det M$ at $k_2/k_1=$} \\
$J$ & $N$ & pts & doublets & $0.05$ & $0.3$ & $0.9$ \\
\hline
$1$ & $3$ & $4$ & $4$ & $-4.5\!\times\!10^{-2}$ & $-1.8\!\times\!10^{-1}$ & $-8.3\!\times\!10^{-2}$ \\
$3/2$ & $4$ & $9$ & $10$ & $-2.2\!\times\!10^{-3}$ & $-9.3\!\times\!10^{-2}$ & $-9.2\!\times\!10^{-3}$ \\
$2$ & $5$ & $16$ & $20$ & $-1.2\!\times\!10^{-4}$ & $-3.3\!\times\!10^{-2}$ & $-9.9\!\times\!10^{-4}$ \\
$5/2$ & $6$ & $25$ & $35$ & $-3.0\!\times\!10^{-6}$ & $-6.6\!\times\!10^{-3}$ & $-5.3\!\times\!10^{-5}$ \\
$3$ & $7$ & $36$ & $56$ & $-9.5\!\times\!10^{-8}$ & $-1.4\!\times\!10^{-3}$ & $-3.4\!\times\!10^{-6}$ \\
$7/2$ & $8$ & $49$ & $84$ & $-2.0\!\times\!10^{-9}$ & $-2.1\!\times\!10^{-4}$ & $-1.5\!\times\!10^{-7}$ \\
$4$ & $9$ & $64$ & $120$ & $-4.8\!\times\!10^{-11}$ & $-3.5\!\times\!10^{-5}$ & $-7.4\!\times\!10^{-9}$
\end{tabular}
\end{ruledtabular}
\end{table}

In every case the commutators of $H_{mn}$ with $P_m$ and $Q_n$ vanish to
machine precision, with a largest Frobenius norm of $5.5\times10^{-14}$ over
the whole sweep. The observed number of doublets equals $f(m,n)$, no eigenvalue
has multiplicity above two, the ranks match \eqref{eq:rank}, and $\det M<0$ for
every doublet. The factorized formula \eqref{eq:detM} reproduces the directly
computed determinant to a relative error below $10^{-9}$. The residual is
round-off: at four lattice points, among them the one with the largest
residual, the two agree to all digits in $60$-digit arithmetic~\cite{supp}.

\begin{figure}[tbp]
\centering
\includegraphics[width=0.92\columnwidth]{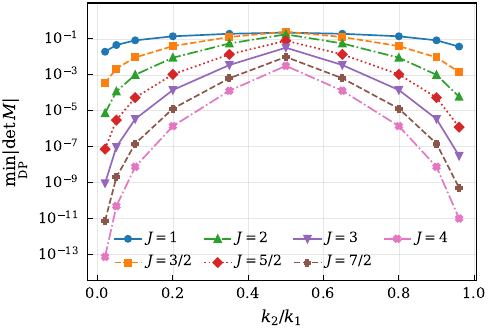}
\caption{Smallest $|\det M|$ over the entire DP lattice as a function of the
anisotropy ratio, for $J=1,\tfrac32,\dots,4$. The determinant is strictly
negative everywhere (Result~\ref{res:chern}). Its modulus falls by many orders
of magnitude toward either uniaxial limit, because the product $\alpha\beta$ of
the boundary amplitudes in Eq.~\eqref{eq:detM} tends to zero there for some
doublets (Sec.~\ref{sec:chern}); a small $|\det M|$
means a cone whose splitting rises slowly in at least one field direction,
with unchanged orientation. The minimum decreases with $J$ at every ratio
except in a narrow window around $k_2/k_1=\tfrac12$, where the $J=\tfrac32$
curve lies slightly above the $J=1$ one.}
\label{fig:detM}
\end{figure}

Figure~\ref{fig:detM} shows the smallest $|\det M|$ over the whole lattice.
At $J=3$ and $k_2/k_1=0.05$ it is $9.5\times10^{-8}$, and at $J=4$ it is
$4.8\times10^{-11}$.

\begin{figure}[tbp]
\centering
\includegraphics[width=0.92\columnwidth]{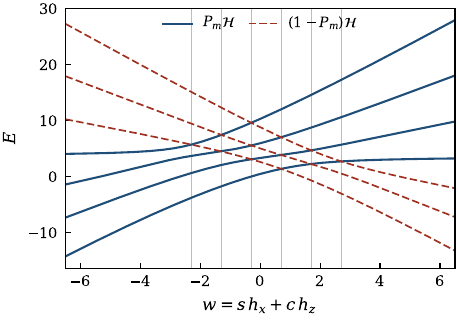}
\caption{Spectrum along the cut line $m=0$ for $J=3$ and $k_2/k_1=0.3$,
parametrized by $w=sh_x+ch_z$; $E$ and $w$ are in units of $k_1$. The severed bond makes the spectrum the union of
the two segments' spectra (solid and dashed), and levels within one segment
never cross. Each of the $p(N-p)=12$ pairs drawn from the two segments crosses
exactly once, and all crossings sit at the $2J=6$ lattice points of the line
(vertical rules).}
\label{fig:cutline}
\end{figure}

Figure~\ref{fig:cutline} shows the spectrum along one cut line.

% =====================================================================
\section{Beyond the quadratic model}
\label{sec:beyond}
% =====================================================================

This section leaves the exactly solvable model. The statements on the
commutator, on the continuation of each doublet for small $|C|$, on the
Kramers doublets at zero field, on charge conservation and on the surviving
symmetries are proven; the separation of the branches that leave a
multiple point and all values computed for Fe$_8$ are numerical observations. Everything
above rests on nearest-neighbor hopping in each rotated frame, that is, on the
vanishing of the $X_\varepsilon^2$ coefficient $k_1c^2-k_2$
(Appendix~\ref{app:chain}). A fourth-order term $-C(S_+^4+S_-^4)$, written
with the sign and the value $C=29\,\mu$K of Ref.~\cite{LiGarg}, is present in
Fe$_8$ and is required for a quantitative account of its spectrum. It adds hoppings of range up to four, a vanishing nearest-neighbor amplitude no
longer disconnects the chain, and the cut projectors cease to commute with
$H$. Since the quadratic model is tridiagonal in the $\Zp$ basis at every
field, the $(r{+}4,r)$ element of $[H,P_m]$ for $r\le m<r+4$ is
$-C\me{r{+}4}{S_+^4+S_-^4}{r}=-C(\cos^8\tfrac\theta2+\sin^8\tfrac\theta2)\,
\ell_r\ell_{r+1}\ell_{r+2}\ell_{r+3}\neq0$. For $J\ge2$ such an element exists
for every $m$, so for $C\neq0$ no $P_m$, and likewise no $Q_n$, commutes with
$H$ at any field. For
sufficiently small $|C|$ each doublet, being spectrally isolated with
$\det M\neq0$, continues as a smooth branch of DPs by the implicit function
theorem. The theorem does not say whether the $f$
branches leaving a multiple point separate. Away from zero field they separate in every case computed. For
$J=2,\tfrac52,\dots,4$ and the three anisotropy ratios of
Sec.~\ref{sec:numerics}, the first-order velocities of the $f$ branches are
pairwise distinct at every lattice point with $f(m,n)\ge2$ and $\bh\neq0$. At $J=3$ and $k_2/k_1=0.3$, for example, the three doublets
at $(m,n)=(0,-1)$, at $h_x=k_1s$, move apart along the $h_x$ axis with a spread
that grows linearly in $C$ and reaches $1.5\times10^{-2}k_1$ at
$C=10^{-4}k_1$.

At zero field they do not separate. For half-integer $J$ the $N/2$ doublets
at $\bh=0$ (Sec.~\ref{sec:counting}\,B) are Kramers doublets, and
$S_+^4+S_-^4$ is even under time reversal,
so the perturbed Hamiltonian at zero field is still time-reversal invariant
and every level stays Kramers degenerate. All $N/2$ doublets therefore remain
at $\bh=0$ for any $C$.

At $C=0$ every DP is a monopole of charge $-1$ (Result~\ref{res:chern}); since
the total charge is conserved, a DP can move under the perturbation but can
disappear only by meeting one of opposite charge. The quartic term is invariant under the $\pi$ rotations about $x$, $y$ and
$z$ and under time reversal, so every symmetry of the family $H(\bh)$ survives
it; only the cuts are lost. Complex conjugation $\mathcal K$ in the $S_z$ basis maps
$H(h_x,h_y,h_z)$ to $H(h_x,-h_y,h_z)$, and $\mathcal KR_x(\pi)$, with
$R_x(\pi)=e^{-\ii\pi S_x}$, maps it to $H(h_x,h_y,-h_z)$. Both are antiunitary
and square to one, so on each of the planes $h_y=0$ and $h_z=0$ the
Hamiltonian is real in a suitable basis, the codimension is two, and a DP can
leave either plane only together with its mirror image. On the $h_x$ and $h_z$
axes the $\pi$ rotation about the field sorts the states into two sectors, and
a crossing between them cannot be lifted; it can disappear only by meeting
another crossing of the same two levels. In a real basis of the two sector
states $M$ has at most one nonzero entry per row and column,
$\pm\tfrac12\partial(E_+-E_-)/\partial h$, whose sign alternates from one
crossing to the next, and a matrix element between the sectors of each
transverse spin component. Crossings that merge while these elements are
nonzero thus carry opposite charges and annihilate. Since crossings continued
from $C=0$ carry $\chi=-1$, one of the two first reverses its charge where a
transverse element vanishes, emitting a mirror pair of DPs off the axis. For
the two lowest levels of Fe$_8$ (parameters below) this happens on the
positive hard axis near $C=2.43$, $4.99$ and $11.91\,\mu$K, the pair entering
the plane $h_z=0$ less than $0.01\,\mu$K before the annihilation.

Counting a degeneracy at every field $h>0$ for every pair of levels that meets
there, so that a point of multiplicity $f$ counts $f$ times, a census at $J=10$
with the Fe$_8$ values $k_1=0.338$~K, $k_2=0.246$~K and
$C=29\,\mu$K~\cite{LiGarg} gives $55$ on the positive $h_z$ half-axis, the
number found at $C=0$, but only $31$ on the positive $h_x$ half-axis, where
$C=0$ also gives $55$. The sum rules constrain only the signed index summed
over all of field space and therefore allow this. Bruno argued
from them that the missing points cannot disappear and predicted that those of
the two lowest levels collide pairwise and leave the hard axis into the
$h_x$--$h_y$ plane~\cite{Bruno}; Li and Garg~\cite{LiGarg} located them
numerically.
Earlier work had explained the absence of quenches beyond the fourth by a
change of the dominant instanton above a critical field, where an instanton
with boundary jumps and no interfering partner takes
over~\cite{KG2002,KG2003}.

In the same census, of the ten DPs between the two lowest levels (the pair
whose hard-axis quenches were observed~\cite{WernsdorferSessoli}) that the
quadratic model puts on the positive hard axis, four survive the fourth-order
term on the axis, at $h_x=0.273$, $0.816$, $1.359$ and $1.734$~K, and the other
six are replaced by three mirror pairs in the plane $h_z=0$, at
$(h_x,h_y)=(2.122,\pm0.434)$, $(2.620,\pm0.901)$ and $(3.148,\pm1.381)$~K.
Before rounding to the three decimals given here, both sets reproduce, in
units of $2k_1J=6.76$~K, the positions tabulated in
Ref.~\cite{LiGarg} to all four decimals quoted there, and the three pairs are
the three points predicted in Ref.~\cite{Bruno} on each of the two branches
that leave the hard axis.

% =====================================================================
\section{Conclusions}
\label{sec:conclusions}
% =====================================================================

At each point of the Ke\c{c}ecio\u{g}lu--Garg lattice the two cuts of the
biaxial spin chain are symmetries of the Hamiltonian, explicit before any
diagonalization and not commuting with each other. Their noncommutation forces
the degeneracies, the rank of the pairing operator counts them, and the
disjoint supports of the doublet partners fix the orientation of every cone.
For $J\ge2$ the fourth-order term $-C(S_+^4+S_-^4)$ destroys the cuts, but
for small $|C|$ the DPs persist with unit charge (Sec.~\ref{sec:beyond}).

\section*{Data availability}

The data that support the findings of this article are openly
available~\cite{code}. All numerical data underlying the figures and tables
are generated by a self-contained script that uses only standard numerical
libraries. When run with the \texttt{-{}-verify} option, it also recomputes the
quantitative claims of Sec.~\ref{sec:numerics}, Table~\ref{tab:summary} and
Table~S1 of the Supplemental Material, all stated thresholds and the values
computed for Fe$_8$ in Sec.~\ref{sec:beyond}, compares them with the manuscript
source (the census with reference values stored in the script), and exits with
an error on any discrepancy.

\section*{Use of artificial-intelligence tools}

Generative artificial-intelligence tools were used in this work and are
disclosed here as required by APS policy. ChatGPT (OpenAI, GPT-5.6
``Sol'') assisted in developing the two-cut interpretation and initial
derivations. Claude Opus 5 (Anthropic) assisted with proof development,
manuscript editing and the verification script, and Claude Opus 5.5
(Anthropic) with the revision in response to the referees. ChatGPT, Claude
Opus 5, Claude Opus 5.5 and Kimi K3 (Moonshot AI) were also used for
independent attempts to reproduce or refute algebraic and numerical claims.
The author directed these uses, independently checked the resulting equations
and numerical calculations, rejected unsupported output, and accepts full
responsibility for the manuscript.

\begin{acknowledgments}
S.G.T. was supported by the Deutsche Forschungsgemeinschaft (DFG) under
Project 535298924.
\end{acknowledgments}

\appendix

% =====================================================================
\section{The chain, and why the lattice is exhaustive}
\label{app:chain}
% =====================================================================

Rotating the frame about $y$ through $\varepsilon\theta$, with $\theta$ not yet
fixed, so that $S_x=\cos\theta\,X_\varepsilon+\varepsilon\sin\theta\,Z_\varepsilon$
and $S_z=-\varepsilon\sin\theta\,X_\varepsilon+\cos\theta\,Z_\varepsilon$, and
eliminating $Y^2$ with $X_\varepsilon^2+Y^2+Z_\varepsilon^2=J(J{+}1)\one$, one
finds
\begin{equation}
\begin{aligned}
  k_1S_x^2+k_2S_y^2={}&(k_1\cos^2\theta-k_2)X_\varepsilon^2
   +(k_1\sin^2\theta-k_2)Z_\varepsilon^2\\
  &+\varepsilon k_1\cos\theta\sin\theta\{X_\varepsilon,Z_\varepsilon\}
   +k_2J(J{+}1)\one .
\end{aligned}
  \label{eq:expand}
\end{equation}
In the $Z_\varepsilon$ basis $X_\varepsilon^2$ is the only term with nonzero
$(r{+}2,r)$ elements, $\me{r{+}2}{X_\varepsilon^2}{r}=\ell_r\ell_{r+1}/4$, so
for $J\ge1$ the Hamiltonian is tridiagonal exactly when $k_1\cos^2\theta=k_2$.
For $0<k_2<k_1$ this fixes $\cos^2\theta=c^2$. Since $\theta\to\theta+\pi$
only reverses the rotated axes, $(X_\varepsilon,Z_\varepsilon)\to
(-X_\varepsilon,-Z_\varepsilon)$, and $\theta\to-\theta$ is the same as
$\varepsilon\to-\varepsilon$, we may take $\cos\theta=c$, $\sin\theta=s$,
$0<\theta<\pi/2$; the two values $\varepsilon=\pm1$ then give the two frames,
and the other two coefficients become $k_1s^2-k_2=k_1-2k_2$ and
$\varepsilon k_1cs=\varepsilon a$, which is Eq.~\eqref{eq:Hrot}. The matrix elements behind
Eq.~\eqref{eq:hop} are listed in the Supplemental Material~\cite{supp}. For
$J=\tfrac12$ every $2\times2$ matrix is tridiagonal and the condition is
empty, but the same choice of $\theta$ is admissible and all formulas apply:
the anisotropy is $\tfrac14(k_1+k_2)\one$, the only lattice point is
$m=n=-\tfrac12$ at $\bh=0$ with $f=1$, and Eq.~\eqref{eq:detM} gives
$\det M=-\tfrac18$, the value for $-\bh\cdot\bS$. In either uniaxial limit the
two eigenbases agree up to the ordering of the sites (Sec.~\ref{sec:chain}),
and the construction fails there, as it does when a higher-order anisotropy is
added.

In a chain with no vanishing hopping, fix an eigenvalue: the three-term
recurrence determines every site amplitude from the first one, so the
eigenvalue cannot be twice degenerate. If the amplitude on the first site
vanished, the first row of the eigenvalue equation would force the amplitude on the
second site to vanish, and induction would annihilate the whole state, so the amplitude at
either end site is never zero.

Simplicity of the unsevered chain puts the DPs on a lattice. If $H(\bh)$ is
degenerate, some hopping \eqref{eq:hop} of the $\varepsilon=+1$ chain must
vanish. Since $\ell_r>0$ for $-J\le r\le J-1$, both the real and the imaginary
part of the bracket must vanish, i.e.
\begin{equation}
  h_y=0,\qquad ch_x-sh_z=a(2m+1)
  \label{eq:cut1}
\end{equation}
for a unique $m$, and the $\varepsilon=-1$ chain gives, for a unique $n$,
$h_y=0$ and $ch_x+sh_z=-a(2n+1)$. Adding and subtracting the two conditions and
using $a=k_1cs$ yields \eqref{eq:lattice}, while $-J\le r\le J-1$ gives the
range \eqref{eq:mnrange}. Since $(m,n)\mapsto(m-n,m+n)$ is injective, distinct
index pairs give distinct fields. The dictionary between these labels and
those of Ref.~\cite{KG} is in the Supplemental Material~\cite{supp}.

% =====================================================================
\section{Star counting and the rank of the pairing operator}
\label{app:stars}
% =====================================================================

\emph{Star content of the cut subspaces.} In the stellar
representation~\cite{Majorana} a state $\psi\neq0$ is represented by its $2J$
Majorana stars, counted with multiplicity: the antipodes of the zeros of the
coherent-state amplitude $\ip{\hat n}{\psi}$ as a function of $\hat n$, as made
precise in the next paragraph. A state with all $2J$ stars at $\hat n$ is the coherent
state along $\hat n$, and the basis state $\ket{r}$ quantized along $\hat n$
has $J+r$ stars at $\hat n$ and $J-r$ at $-\hat n$. The range of $P_m$ is
spanned by the $\ket{r;+}$ with $r\le m$, every one of which carries at least
$J-m$ stars at $-\hat n_+$, where $\hat n_\pm=(\pm s,0,c)$ are the two rotated
Bloch directions. Carrying at least $\xi$ stars at a given point is a linear
condition, so the states satisfying it form a subspace, of dimension
$2J-\xi+1$ by the count in the next paragraph (take $\eta=0$ there). For
$\xi=J-m$ at $-\hat n_+$ this dimension is $J+m+1=p$; the subspace contains
$\Ran P_m$, which also has dimension $p$, so the two coincide. The other three
lines of \eqref{eq:starspaces} follow in the same way. Altogether
\begin{align}
  \Ran P_m&:\ \text{at least } J-m \text{ stars at } -\hat n_+, \notag\\
  \ker P_m&:\ \text{at least } p \text{ stars at } +\hat n_+, \notag\\
  \Ran Q_n&:\ \text{at least } J-n \text{ stars at } -\hat n_-, \notag\\
  \ker Q_n&:\ \text{at least } q \text{ stars at } +\hat n_- .
  \label{eq:starspaces}
\end{align}

\emph{Two clusters.} For the coherent state
$\ket\zeta\propto e^{\zeta S_-}\ket J$ along the direction with stereographic
coordinate $\zeta$, the amplitude is $\ip{\zeta}{\psi}\propto\Phi_\psi(\bar\zeta)$
with the Majorana polynomial $\Phi_\psi(\omega)=\me{J}{e^{\omega S_+}}{\psi}$, whose
coefficient of $\omega^{J-r}$ is a nonzero multiple of $\ip{r}{\psi}$. Thus
$\psi\mapsto\Phi_\psi$ is a linear bijection onto the polynomials of degree at
most $2J$, and the amplitude vanishes at $-\hat z$ to the order
$2J-\deg\Phi_\psi$. The four directions $\pm\hat n_+$, $\pm\hat n_-$ are distinct and differ from
$\pm\hat z$ because $0<s,c<1$. A star of multiplicity $\xi$ at
$\hat n\neq\hat z$ is a zero of $\Phi_\psi$ of order $\xi$ at the finite point
$\omega=\bar\zeta$, where $\zeta$ is the stereographic coordinate of $-\hat n$.
A demand for $\xi$ stars at one of the four directions and $\eta$ stars at
another, $\xi,\eta\in\mathbb Z_{\ge0}$, is therefore divisibility of
$\Phi_\psi$ by $(\omega-\omega_1)^\xi$ and by $(\omega-\omega_2)^\eta$, with the finite points
$\omega_1\neq\omega_2$ so assigned. The two factors are coprime, so the
states meeting both demands are the multiples of one fixed polynomial of
degree $\xi+\eta$. Such multiples exist only for $\xi+\eta\le2J$, and then the
cofactor is an arbitrary polynomial of degree at most $2J-\xi-\eta$, so they
form a space of dimension $2J-\xi-\eta+1$.

\emph{The four common eigenspaces.} Applied to \eqref{eq:starspaces}, this
gives the dimensions of $\HS_{ij}=\{\psi:P_m\psi=i\psi,\ Q_n\psi=j\psi\}$,
\begin{align}
  d_{11}&=(p+q-N)_+, & d_{10}&=(p-q)_+, \notag\\
  d_{01}&=(q-p)_+, & d_{00}&=(N-p-q)_+ ,
  \label{eq:dij}
\end{align}
with $x_+=\max(x,0)$. For instance, $\HS_{11}$ demands $J-m$ stars at
$-\hat n_+$ and $J-n$ at $-\hat n_-$, which requires $(J-m)+(J-n)\le2J$ and
then leaves $m+n$ free stars, i.e.\ dimension $m+n+1=p+q-N$. These are the
smallest dimensions compatible with the ranks $p$ and $q$; in this sense the
two cuts are in general position. Since $d_{11}>0$ requires $p+q>N$ while
$d_{00}>0$ requires $p+q<N$, at most one of $\HS_{11},\HS_{00}$ is nonzero;
likewise, since $d_{10}>0$ requires $p>q$ and $d_{01}>0$ requires $q>p$, at
most one of $\HS_{10},\HS_{01}$ is nonzero.

\emph{Rank of the pairing operator.} $B_{mn}=P_mQ_n(1-P_m)$ annihilates the
lower segment $P_m\HS$. For $y$ in the upper segment $(1-P_m)\HS$, $P_mQ_ny=0$
says that $Q_ny$ lies in the upper segment. Writing $y=(y-Q_ny)+Q_ny$, the
first piece is annihilated by $Q_n$ and the second is fixed by it, and both
lie in the upper segment. The kernel of $B_{mn}$ inside the upper segment is
therefore $\HS_{01}\oplus\HS_{00}$, and
\begin{align}
  \rank B_{mn}&=(N-p)-d_{01}-d_{00} \notag\\
              &=\min\{p,q,N-p,N-q\}=f(m,n) ,
\end{align}
where the second line is \eqref{eq:dij} evaluated in each of the four cases fixed by the signs of $p-q$ and
$p+q-N$. With \eqref{eq:commB} this also gives
$\rank[P_m,Q_n]=2f(m,n)$, and $1\le p,q\le N-1$ gives $f(m,n)\ge1$, so every
lattice point carries at least one degeneracy.

The four sectors $\HS_{11},\HS_{10},\HS_{01},\HS_{00}$ and their orthogonal
complement $\Ran[P_m,Q_n]=\Hpair$ form the canonical decomposition of a pair of
orthogonal projections~\cite{Halmos}; on the complement the two projections are
in what Halmos calls generic position. The general position of
Result~\ref{res:transversal} only asks that the four sectors be as small as the
ranks allow; the star count supplies their dimensions.

% =====================================================================
\section{The level-pair count}
\label{app:levels}
% =====================================================================

Number the global levels $1,\dots,N$ from bottom to top, fix a cut $m$ and move
along the line \eqref{eq:cut1}, on which $u_{+1}$ and $v$ are held fixed and
$w_{+1}$ is free; the subscript is dropped in the rest of this appendix. On the line, Eq.~\eqref{eq:hop} gives $t^{(+1)}_r=a\ell_r(r-m)$, independently
of $w$: the bond $m\leftrightarrow m+1$ is severed along the whole line, and
every other hopping is nonzero there. The spectrum is therefore the union of
the two segments' spectra, each of them simple (Appendix~\ref{app:chain}). Label them
$E^L_1<\dots<E^L_p$ and $E^R_1<\dots<E^R_{N-p}$. Every eigenstate of the lower
segment is supported on the sites $r\le m$ and every eigenstate of the upper
segment on the sites $r\ge m+1$, so $\langle\Zp\rangle_{L,i}\le m<m+1\le
\langle\Zp\rangle_{R,j}$ for every $i$ and $j$, which is \eqref{eq:zgap}
extended from the doublet partners to all levels. Since $H$ depends on $w$ only
through $-w\Zp$, the Hellmann--Feynman theorem gives
\begin{equation}
  \frac{d}{dw}\bigl(E^L_i-E^R_j\bigr)
   =-\langle \Zp\rangle_{L,i}+\langle \Zp\rangle_{R,j}\ \ge\ 1 ,
  \label{eq:HF}
\end{equation}
so every left--right level difference increases strictly, with slope at least
one, from $-\infty$ to $+\infty$. Each of the $p(N-p)$ pairs therefore crosses
exactly once. A crossing is a degeneracy of $H$, so the $\varepsilon=-1$ chain
is severed there as well and \eqref{eq:lattice} applies: every crossing sits at
a lattice point of the line. Levels within one segment never cross
(Fig.~\ref{fig:cutline}). The distribution of the $p(N-p)$ crossings over the
$2J$ points is derived below; their number at $(m,n)$ is $f(m,n)$, which is
\eqref{eq:rowsum}.

At a crossing $E^L_i=E^R_j$ there are $(i-1)+(j-1)$ levels below it, so it joins
the global levels $k=i+j-1$ and $k+1$. Summing over the cuts, write $k+1=i+j$:
the requirement $j\ge1$ leaves $k$ choices of $i$, and for each of them
$j\le N-p$ and $i\le p$ leave the cuts with $i\le p\le N-k-1+i$, that is
$N-k$ of them. Hence $N_k=k(N-k)$, and
$\sum_{k=1}^{N-1}k(N-k)=\tfrac16N(N^2-1)=\tfrac23J(J+1)(2J+1)$ is the total
number of crossings, which by the same identification is \eqref{eq:total}.

\emph{Position of each crossing.} The same monotonicity fixes the position of
each crossing, Eq.~\eqref{eq:whichpoint}. Write $w_{ij}$ for the field at
which $E^L_i=E^R_j$. Since $E^L_{i+1}>E^L_i$ at every $w$, the difference
$E^L_{i+1}-E^R_j$ exceeds $E^L_i-E^R_j$ everywhere, and as both increase
strictly through zero by \eqref{eq:HF}, the zero of the larger lies to the
left: $w_{i+1,j}<w_{ij}$, and likewise $w_{i,j+1}>w_{ij}$. The available
positions are the $N-1$ lattice points of the line, at
$w^{(m,n)}=k_1[s^2(m-n)-c^2(m+n+1)]$, which is strictly decreasing in $n$.
Let $\phi(i,j)=J+n_{ij}+1\in\{1,\dots,N-1\}$ number the point at which the pair
$(i,j)$ crosses. Then $\phi$ increases strictly with $i$ and decreases strictly
with $j$. Walking from $(1,N-p)$ to $(i,j)$ in $(N-p-j)+(i-1)$ steps, each
raising $\phi$ by at least one, gives $\phi\ge N-p+i-j$; continuing to $(p,1)$
and using $\phi\le N-1$ gives $\phi\le N-p+i-j$. Hence $\phi(i,j)=N-p+i-j$,
and with $N-p=J-m$ this is \eqref{eq:whichpoint}. The two bounds meet because a
maximal chain in the $(i,j)$ ordering has length $N-1$, the number of
available positions.

The formula has three consequences. The crossing $i=j=1$ gives $n=-m-1$ and
hence \eqref{eq:groundDP}, and since \eqref{eq:Nk} gives $N_1=2J$ DPs between
the two lowest levels, the $2J$ cut lines account for all of them: no DP
between the two lowest levels lies off the hard axis, at any $J$ and any
anisotropy. The crossings at one lattice point are those with $i-j$ fixed at
$m+n+1$, and the number of pairs with $1\le i\le p$, $1\le j\le N-p$ and
$i-j=p+q-N$ is $\min\{p,q,N-p,N-q\}=f(m,n)$, so the multiplicity \eqref{eq:f}
follows from the chain alone, independently of the star count of
Appendix~\ref{app:stars}. Finally, since the crossing $(i,j)$ joins the global
levels $k=i+j-1$ and $k+1$, the DPs at which levels $k$ and $k+1$ meet form the
rectangular array $X=-(N-k-1),-(N-k-3),\dots,N-k-1$ and
$Z=-(k-1),-(k-3),\dots,k-1$, of size $k(N-k)$: for all levels, the
semiclassical assignment~\cite[Eqs.~(2) and (3)]{KG}, whose well indices
$\ell',\ell''$ give the global pair $k=\ell'+\ell''+1$, $k+1$, obtained here
from the chain.

\end{document}

\title{Supplemental Material:\texorpdfstring{\\}{ }
Hidden symmetry at the diabolical points of a biaxial spin}

\author{Shadan Ghassemi Tabrizi}
\email{s.ghassemi-tabrizi@hzdr.de}
\affiliation{Computational System Sciences, Technische Universit\"at Dresden,
01187 Dresden, Germany}
\affiliation{Center for Advanced Systems Understanding (CASUS),
Am Untermarkt 20, 02826 G\"orlitz, Germany}

\date{November 11, 2026}

\maketitle

% Equations of this document carry an S, so that they cannot be confused
% with the article's, which are cited by bare number throughout.
\renewcommand{\theequation}{S\arabic{equation}}
\renewcommand{\thesection}{S\arabic{section}}
\renewcommand{\thetable}{S\arabic{table}}

This document contains material that supports the article without being
needed for any of its four Results: the algebra behind
Eqs.~(7) and (12), the dictionary between the lattice
labels used here and those of Ke\c{c}ecio\u{g}lu and Garg~\cite{KG}, a second
proof of the transversality of the two cuts by way of a determinant identity,
with a closed lower bound on the singular values of the pairing operator, the
explicit form of the pseudospin, the numerical method with all of its
thresholds, and the complete point-by-point data for $J=2$ at $k_2/k_1=0.3$. Equations, sections and the table of this
document carry the prefix S, as in (S1), Sec.~S1 and Table~S1; unless stated
otherwise, all other equation, section, appendix, figure and Result numbers
refer to the article. The notation is that
of the article.

% =====================================================================
\section{Algebra of the two rotated frames}
% =====================================================================

\emph{The two-axis form.} With $c^2=k_2/k_1$ and $s^2=1-k_2/k_1$ one has
$S_x=(\Zp-\Zm)/(2s)$ and $S_z=(\Zp+\Zm)/(2c)$, and since
$\{\Zp,\Zm\}=2c^2S_z^2-2s^2S_x^2$ together with
$S_x^2+S_y^2+S_z^2=J(J{+}1)\one$,
\begin{equation}
  k_1S_x^2+k_2S_y^2=k_2J(J{+}1)\one-\frac{k_1}{2}\{\Zp,\Zm\} .
  \label{eq:quadid}
\end{equation}
When the lattice field, Eq.~(8) of the article, is inserted into the Zeeman term,
the coefficient of $\Zp$ becomes $\tfrac{k_1}{2}[-(m-n)+(m+n+1)]=k_1(n+\tfrac12)$
and that of $\Zm$ becomes $k_1(m+\tfrac12)$; together with \eqref{eq:quadid}
this gives Eq.~(12) of the article.

\emph{The chain form.} The expansion of the anisotropy in the rotated frame is
Eq.~(A1) of the article; at $\cos\theta=c$ its $X_\varepsilon^2$ term vanishes. In the $Z_\varepsilon$ eigenbasis, in the phase convention
fixed in Sec.~II of the article, only $X_\varepsilon$, $Y$ and
$\{X_\varepsilon,Z_\varepsilon\}$ are off-diagonal, with
$\me{r{+}1}{X_\varepsilon}{r}=\ell_r/2$, $\me{r{+}1}{Y}{r}=-\ii\ell_r/2$ and
$\{X_\varepsilon,Z_\varepsilon\}_{r+1,r}=(2r+1)\ell_r/2$, which gives
Eq.~(7) of the article.

\emph{Uniqueness of the algebraic Heun representative.} A Hermitian operator
affine in each member of the pair is
$\lambda\{\Zp,\Zm\}+\lambda'\,\ii[\Zp,\Zm]+\lambda_+\Zp+\lambda_-\Zm+\lambda_0\one$,
and $\ii[\Zp,\Zm]=2csS_y$, so the antisymmetric part is a field along $y$. In
the $\Zp$ basis neither the diagonal $\Zp$ term nor the constant contributes to
the hopping across $r\leftrightarrow r+1$; collecting the other three gives
$-cs\,\ell_r[\lambda(2r+1)+\ii\lambda'+\lambda_-]$. Commuting with $P_m$ makes
this vanish at $r=m$, and since real and imaginary parts must vanish
separately, $\lambda'=0$ and $\lambda_-=-\lambda(2m+1)$. The dual condition in
the $\Zm$ basis gives $\lambda_+=-\lambda(2n+1)$. Only $\lambda$ and
$\lambda_0$ are left, and $\lambda=-k_1/2$, $\lambda_0=k_2J(J+1)$ give
Eq.~(12) of the article.

% =====================================================================
\section{Index conventions of Ref.~\texorpdfstring{\cite{KG}}{[1]}}
% =====================================================================

Reference~\cite{KG} writes the model as its Eq.~(1),
$-k_2J_z^2+(k_1-k_2)J_x^2-g\mu_B\mathbf J\cdot\mathbf H$, which is
Eq.~(1) of the article up to a constant, and uses scaled fields $u_x$ and $u_z$ that are
the scaled coordinates $X$ and $Z$ of Sec.~II of the
article. Its first rotation gives the cut lines $u_x+u_z=-(2m_0+1)$, its
Eq.~(9), the mirrored rotation $u_x-u_z=-(2n_0+1)$, its Eq.~(10), with
$m_0,n_0\in\{-J,\dots,J-1\}$, and the lattice is its Eq.~(11),
$u_x=-(m_0+n_0+1)$, $u_z=n_0-m_0$. The dictionary
\begin{equation}
  m=-n_0-1,\qquad n=m_0
  \label{eq:dict}
\end{equation}
gives $X=m-n=u_x$, $Z=-(m+n+1)=u_z$, $|m-n|=|m_0+n_0+1|$ and
$|m+n+1|=|m_0-n_0|$, so that Eq.~(16) of the article and Eq.~(17) of Ref.~\cite{KG} agree
term by term; both index sets run over $\{-J,\dots,J-1\}$. The first rotation
of Ref.~\cite{KG} is the $\varepsilon=-1$ frame of the article, whose cut at
$n$ is the line $X+Z=-2n-1$, and its two blocks have sizes $J+m_0+1=q$ and
$J-m_0=N-q$, the ranks of $Q_n$ and $1-Q_n$; the blocks of the mirrored
rotation have sizes $p$ and $N-p$, the ranks of $P_m$ and $1-P_m$. Reference~\cite{KG} obtains the multiplicities by deforming the anisotropy to
an end point at which the rotated Hamiltonian is diagonal and counting the
solutions of a condition of the form $m_1+m_2=\text{const}$, its
Eqs.~(15)--(17). The rotated Hamiltonian is diagonal at both end points: at
$k_2/k_1=0$ the condition reads $m_1+m_2=m_0+n_0+1$, and at $k_2/k_1=1$ it
reads $m_1+m_2=m_0-n_0$, the form of Eq.~(16) of Ref.~\cite{KG}, which that
work states for $k_2/k_1=0$; both give the count of its Eq.~(17). The combination $m_0-n_0=m+n+1$ is the one that appears
in Eq.~(22) of the article, which Appendix~C of the article
derives at the physical anisotropy.

% =====================================================================
\section{Transversality from a determinant identity}
\label{sec:det}
% =====================================================================

The four dimensions of the common eigenspaces of $P_m$ and $Q_n$ are obtained in
Appendix~B of the article by counting Majorana stars. They also
follow from a determinant identity, which in addition gives quantitative
bounds. The overlap matrix of the two bases is the Wigner rotation matrix
$d^J(2\theta)$, up to the sign of the angle. The Gauss decomposition in
$\mathrm{SL}(2,\mathbb C)$, applied to that SU(2) element and carried into the
spin-$J$ representation, writes it as
$d^J(2\theta)=e^{-\tan\theta\,S_+}\,c^{-2S_z}\,e^{\tan\theta\,S_-}$, which in
the ordering $r=-J,\dots,J$ is a lower triangular times a diagonal times an
upper triangular factor; the factors are not unitary, and the decomposition
exists because $c=\cos\theta\neq0$. The leading $\kappa\times\kappa$ block of
such a product is the product of the corresponding blocks, so its determinant
is $\prod_{i=0}^{\kappa-1}c^{-2(-J+i)}=c^{\kappa(N-\kappa)}$; the symmetry
$d^J_{-r,-r'}=(-1)^{r-r'}d^J_{rr'}$ carries this to the opposite corner, and
$d^J(\pi-2\theta)=d^J(\pi)\,d^J(2\theta)^{\mathsf T}$, whose diagonal Gauss
factor is $s^{-2S_z}$, carries it to the two anticorners. The four
$\kappa\times\kappa$ corner minors therefore satisfy
\begin{equation}
  |\det[d^J]_{\text{corner}}|=c^{\,\kappa(N-\kappa)}\ \ \text{or}\ \
  s^{\,\kappa(N-\kappa)},
  \label{eq:corner}
\end{equation}
the first for the two corners on the diagonal and the second for the two
anticorners, and all of them are nonzero because $0<k_2<k_1$. Since a
submatrix of an orthogonal matrix has all singular values at most one while
their product is the modulus of the determinant, \eqref{eq:corner} bounds the
smallest singular value of each corner block from below. The four blocks of the
overlap matrix cut out by $P_m$ and $Q_n$ are rectangular blocks anchored at
the four corners of $d^J$. Each contains the square corner of $d^J$ whose size
is its smaller dimension, and that corner is nonsingular by \eqref{eq:corner},
so each block has full rank. The block with rows in $\Ran P_m$ and columns in
$\Ran Q_n$ represents $\psi\mapsto P_m\psi$ on $\Ran Q_n$, whose kernel is the
common eigenspace $\HS_{01}$ of Appendix~B of the article, so
$d_{01}=q-\min(p,q)=(q-p)_+$; in the same way the blocks
$(\ker P_m,\Ran Q_n)$, $(\Ran P_m,\ker Q_n)$ and $(\ker P_m,\ker Q_n)$ have the
kernels $\HS_{11}$, $\HS_{00}$ and $\HS_{10}$, which gives the other three
dimensions (B2). Interior minors of $d^J$ do vanish at admissible
anisotropies, the smallest example being $d^1_{00}(2\theta)=\cos2\theta$ at
$k_2/k_1=\tfrac12$, so the argument relies on the blocks being anchored at the
corners of $d^J$.

\emph{Lower bound on the singular values of $B_{mn}$.} Let $\phi$ run over the
principal angles between $\Ran Q_n$ and $\Ran P_m$. The nonzero singular
values of $B_{mn}=P_mQ_n(1-P_m)$ are $\cos\phi\sin\phi$ for the angles with
$0<\phi<\pi/2$; the $\cos\phi$ are singular values of the block of the overlap
matrix with rows in $\Ran P_m$ and columns in $\Ran Q_n$, the $\sin\phi$ of the
block with rows in $\ker P_m$ and columns in $\Ran Q_n$. Both are corner
blocks, the first on the diagonal and the second an anticorner, and their
smallest singular values are bounded below by the moduli of the corner minors of sizes
$\kappa=\min(p,q)$ and $\kappa'=\min(N-p,q)$: each block contains the
$\kappa\times\kappa$ or $\kappa'\times\kappa'$ corner of $d^J$, deleting rows
or columns does not increase any singular value, and the smallest singular
value of a square corner is bounded below by the modulus of its determinant. Hence every
nonzero singular value of $B_{mn}$ satisfies
\begin{equation}
  \sigma\ \ge\ c^{\,\kappa(N-\kappa)}\,s^{\,\kappa'(N-\kappa')} .
  \label{eq:sigmabound}
\end{equation}
At $J=\tfrac12$ the bound \eqref{eq:sigmabound} is attained, $\sigma=cs$, but for larger $J$ it can be far from sharp, most of all near the
uniaxial limits: over $1\le J\le3$ and $k_2/k_1\in\{0.05,0.3,0.5,0.9\}$ the
smallest nonzero singular value at a lattice point exceeds the bound at that
point by factors between $1.05$ ($J=1$) and $4\times10^{6}$ ($J=3$,
$k_2/k_1=0.05$). It is therefore not used for
the numerical rank decisions of Sec.~\ref{sec:num}, which rest on the observed
gap between retained and discarded singular values.

% =====================================================================
\section{The pseudospin}
% =====================================================================

Label the doublets $\nu=1,\dots,f(m,n)$ and let $\ket{L_\nu}\in P_m\HS$ and
$\ket{R_\nu}\in(1-P_m)\HS$ be their normalized partners in the two segments.
Since $h_y=0$ makes $H_{mn}$ real symmetric in the $\Zp$ basis, all of these
may be taken real.
The pairing operator annihilates $\ket{L_\nu}$, because
$(1-P_m)\ket{L_\nu}=0$, and maps $\ket{R_\nu}$ to
$b_\nu\ket{L_\nu}$ with $b_\nu=\me{L_\nu}{Q_n}{R_\nu}$, which is nonzero by
step (iii) of the proof of Result~3. Fix the sign of
$\ket{R_\nu}$ so that $b_\nu>0$. Then
\begin{equation}
  \tau_+=\sum_\nu\ket{L_\nu}\bra{R_\nu},\qquad
  \tau_z=\sum_\nu\bigl(\ket{L_\nu}\bra{L_\nu}-\ket{R_\nu}\bra{R_\nu}\bigr)
  \label{eq:pauli}
\end{equation}
together with $\tau_-=\tau_+^\dagger$ obey the Pauli algebra on the paired
subspace and commute with $H_{mn}$, since each doublet is an eigenspace. On
$\Hpair$ one has $B_{mn}=\sum_\nu b_\nu\ket{L_\nu}\bra{R_\nu}$ and
$B_{mn}^\dagger B_{mn}=\sum_\nu b_\nu^2\ket{R_\nu}\bra{R_\nu}$, so $\tau_+$ is
the polar part $B_{mn}(B_{mn}^\dagger B_{mn})^{-1/2}$ of Eq.~(17) of the article, the
inverse taken on $\supp(B_{mn}^\dagger B_{mn})=\mathrm{span}\{\ket{R_\nu}\}$
and $\tau_+$ extended by zero off that support. This restriction is needed
because, by Eq.~(15) of the article, $B_{mn}^\dagger B_{mn}$ has rank $f(m,n)\le\lfloor N/2\rfloor$
and is therefore singular at every lattice point, and its support is the upper
half $\Hpair\cap(1-P_m)\HS$ of the paired subspace, not $\Hpair$ itself.

Since $\ket{L_\nu}$ lies in the $P_m$ segment and $\ket{R_\nu}$ in the other,
$\tau_z=2P_m-1$ on the paired subspace. The pseudospin is therefore generated
by the segment label of the severed $\Zp$ chain together with the operator
that moves between the two segments. Since the $\tau_i$ generate a full
$2\times2$ matrix algebra commuting with $H_{mn}$, the paired subspace
factorizes as in Eq.~(18) of the article, and the eigenvalues of $K_{mn}$ are
the $f(m,n)$ distinct doublet energies.

% =====================================================================
\section{Numerical method}
\label{sec:num}
% =====================================================================

The check of Sec.~VI of the article was carried out by a
self-contained script, openly available as stated in the article. It
builds $S_{x,y,z}$ in the $S_z$ basis, forms $H$ from Eq.~(1) of the article, and obtains
$P_m$ and $Q_n$ from the eigenbases of $Z_\pm$, whose eigenvalues are checked
against $-J,\dots,J$ to eight digits first. Degeneracies are detected from the level spacings (energies in units of
$k_1$), each spacing divided by the larger of the level magnitude and unity, a
pair
counting as degenerate below $10^{-10}$ and as distinct above $10^{-6}$; a
spacing falling between the two would abort the run, and none did. The
splitting Jacobian, Eq.~(24) of the article, is evaluated on the resulting
two-dimensional eigenspaces in the original field frame $(h_x,h_y,h_z)$ rather
than the rotated one, so that the frame independence of $\sgn\det M$ is
tested, and both the sign and the factorized form, Eq.~(27) of the article, are
checked at every doublet. For that comparison the two partners are taken
segment adapted: in the $\Zp$ eigenbasis $H_{mn}$ is block diagonal with
respect to $P_m$, and diagonalizing its two diagonal blocks separately yields
$\ket L$ and $\ket R$ directly, which makes $\alpha$, $\beta$ and
$\langle\Zp\rangle_{L,R}$ well defined; the direct determinant is formed from
eigenvectors of the full $H_{mn}$. Repeating the comparison in $60$-digit
arithmetic, carried with $15$ guard digits, at four lattice points (two at
$J=4$, two at $J=3$, among them the point with the largest double-precision
residual, $J=4$, $k_2/k_1=0.05$, $(m,n)=(0,0)$) gives agreement to all $60$
digits, so the residual quoted in Sec.~VI of the article is
round-off. Ranks are computed
with a singular-value cutoff of $10^{-13}$. Since the nonzero singular values of $B_{mn}$
are of the form $\cos\phi\sin\phi\le\tfrac12$ (Sec.~\ref{sec:det}), and those
of $[P_m,Q_n]=B_{mn}-B_{mn}^\dagger$ are the same, each twice, no singular
value of either exceeds $\tfrac12$, so no scaling of the cutoff is needed; the cutoff has to separate
the two populations, and the run records the ratio of the smallest retained
singular value to it and of it to the largest discarded one, requiring both to
exceed ten. Over the sweep the smallest retained singular value is
$2.0\times10^{-7}$, at $J=4$ and $k_2/k_1=0.05$, and the largest discarded one
$2.2\times10^{-15}$; the worst of the two margins is $45$. The nonzero
singular values of $B_{mn}$ decrease rapidly toward either uniaxial limit: at
$J=4$ the smallest of them falls below $10^{-8}$ for $k_2/k_1<0.024$ and
$k_2/k_1>0.976$, so the older threshold of $10^{-8}$, used in earlier
versions of the verification script, would misclassify ranks there, though not at any of the three ratios of the
sweep.

% =====================================================================
\section{Complete lattice data for \texorpdfstring{$J=2$}{J=2}}
% =====================================================================

Table~\ref{tab:J2} lists all sixteen lattice points.

\begin{table*}[t]
\caption{Complete lattice data for $J=2$, $k_2/k_1=0.3$. $X=m-n$ and
$Z=-(m+n+1)$ are the scaled field coordinates of Fig.~2 of
the article; $p,q$ are the projector ranks; $f$ is the multiplicity predicted
by Eq.~(16) of the article and ``obs.'' the observed number of doublets. Every determinant
is negative, as required by Result~4 of the article. For $f=2$
the two determinants are listed in order of increasing doublet energy.}
\label{tab:J2}
\begin{ruledtabular}
\begin{tabular}{rrrrrrrrrrl}
$m$ & $n$ & $X$ & $Z$ & $p$ & $q$ & $f$ & obs. &
$\rank[P_m,Q_n]$ & $\rank B_{mn}$ & $\det M$ \\
\hline
$+1$ & $+1$ & $+0$ & $-3$ & $4$ & $4$ & $1$ & $1$ & $2$ & $1$ & $-0.1076$ \\
$+0$ & $+1$ & $-1$ & $-2$ & $3$ & $4$ & $1$ & $1$ & $2$ & $1$ & $-0.4049$ \\
$+1$ & $+0$ & $+1$ & $-2$ & $4$ & $3$ & $1$ & $1$ & $2$ & $1$ & $-0.4049$ \\
$-1$ & $+1$ & $-2$ & $-1$ & $2$ & $4$ & $1$ & $1$ & $2$ & $1$ & $-0.5873$ \\
$+0$ & $+0$ & $+0$ & $-1$ & $3$ & $3$ & $2$ & $2$ & $4$ & $2$ & $-0.7498,\;-0.0333$ \\
$+1$ & $-1$ & $+2$ & $-1$ & $4$ & $2$ & $1$ & $1$ & $2$ & $1$ & $-0.5873$ \\
$-2$ & $+1$ & $-3$ & $+0$ & $1$ & $4$ & $1$ & $1$ & $2$ & $1$ & $-0.4277$ \\
$-1$ & $+0$ & $-1$ & $+0$ & $2$ & $3$ & $2$ & $2$ & $4$ & $2$ & $-0.4696,\;-0.3100$ \\
$+0$ & $-1$ & $+1$ & $+0$ & $3$ & $2$ & $2$ & $2$ & $4$ & $2$ & $-0.4696,\;-0.3100$ \\
$+1$ & $-2$ & $+3$ & $+0$ & $4$ & $1$ & $1$ & $1$ & $2$ & $1$ & $-0.4277$ \\
$-2$ & $+0$ & $-2$ & $+1$ & $1$ & $3$ & $1$ & $1$ & $2$ & $1$ & $-0.5873$ \\
$-1$ & $-1$ & $+0$ & $+1$ & $2$ & $2$ & $2$ & $2$ & $4$ & $2$ & $-0.7498,\;-0.0333$ \\
$+0$ & $-2$ & $+2$ & $+1$ & $3$ & $1$ & $1$ & $1$ & $2$ & $1$ & $-0.5873$ \\
$-2$ & $-1$ & $-1$ & $+2$ & $1$ & $2$ & $1$ & $1$ & $2$ & $1$ & $-0.4049$ \\
$-1$ & $-2$ & $+1$ & $+2$ & $2$ & $1$ & $1$ & $1$ & $2$ & $1$ & $-0.4049$ \\
$-2$ & $-2$ & $+0$ & $+3$ & $1$ & $1$ & $1$ & $1$ & $2$ & $1$ & $-0.1076$
\end{tabular}
\end{ruledtabular}
\end{table*}